%% file: main.tex
\documentclass[fleqn,10pt]{wlscirep}
\usepackage[utf8]{inputenc}
\usepackage[T1]{fontenc}

\title{Sensing with Twisted Light: Precision Measurement of Fractional Azimuthal Index to Determine Refractive Index}

\author[1,2,3,$\dagger$]{Christopher Perrella}
\author[1,2,$\dagger$]{Aman Anil Punse}
\author[1,2,4]{Anastasiia Zalogina}
\author[5]{Crispin Szydzik}
\author[1,2,6]{Megan Lim}
\author[2,3,7]{Andreas Boes}
\author[3,5]{Arnan Mitchell}
\author[1,2,6]{Kylie R. Dunning}
\author[1,2,3,8,*]{Kishan Dholakia}
\affil[1]{Centre of Light for Life, School of Biological Sciences, University of Adelaide, Adelaide, SA 5005, Australia}
\affil[2]{Institute for Photonics and Advanced Sensing, University of Adelaide, Adelaide, SA 5005, Australia}
\affil[3]{ARC Centre of Excellence in Optical Microcombs for Breakthrough Science (COMBS)}
\affil[4]{Quantum Materials and Nanophotonics, School of Mathematical and Physical Sciences, University of Technology Sydney, Sydney, NSW 2010, Australia}
\affil[5]{Integrated Photonics and Applications Centre, School of Engineering, RMIT University, Melbourne, VIC 3000, Australia}
\affil[6]{Robinson Research Institute, School of Biomedicine, University of Adelaide, Adelaide, SA 5005, Australia}
\affil[7]{School of Electrical and Mechanical Engineering, University of Adelaide, Adelaide, SA 5005, Australia}
\affil[8]{SUPA, School of Physics and Astronomy, University of St Andrews, St Andrews, KY16 9SS, United Kingdom}

\affil[$\dagger$]{These authors contributed equally to this work}
\affil[*]{kishan.dholakia@adelaide.edu.au}

\begin{abstract}
Light beams possessing orbital angular momentum (OAM) have gained significant interest in areas such as optical manipulation, quantum entanglement, and super-resolved imaging. 
In itself, the OAM for a Laguerre-Gaussian beam is proportional to the azimuthal index of the light field, $l$.
It is in fact continuous in nature and a non-trivial parameter to measure.
The ability to determine $l$ precisely would broaden the use of such beams for new applications.
In this study, we generate Laguerre-Gaussian beams of differing $l$ through mode conversion using microscopic spiral phase plates (SPPs).
The exact value of $l$ imparted for a given incident wavelength is dependant upon the refractive index of the media within which the SPP is immersed. 
Here, we show an ultra-precise approach based on laser speckle to measure the azimuthal index of these generated beams to a precision of $2\,{\times}\,10^{-5}$.
This is an improvement of three orders of magnitude over previous studies.
In turn, this leverages an ultra-precise measurement of the refractive index of the medium surrounding the SPP, with a best measured precision of $6.4\,{\times}\,10^{-7}$\,refractive index units.
This is confirmed to be at the shot-noise limit of the system.
Our study interrogates samples of sucrose and haemoglobin, only 300\,pL in volume, within a microfluidic channel. 
This demonstration of an original form of microfluidic refractive index sensor, based on mode conversion to light fields with OAM,  may be multiplexed to measure spatio-temporal variations and gradients within biological samples.

\end{abstract}

\begin{document}

\flushbottom
\maketitle

\thispagestyle{empty}

\section*{Main}
Light fields possessing orbital angular momentum have become a mainstay for a diverse range of key research areas in the last few decades~\cite{allen1992orbital, cheng2025metrology, yao2011orbital}. The fundamental properties and evolution of such beams is also topic of major interest~\cite{allen_iv_1999, rosales-guzman_review_2018}. 
Such fields may have a phase singularity (vortex) embedded within the field. 
The orbital component of the angular momentum arises from inclined wavefronts that lead to a spiralling motion of the energy flow (Poynting vector) of the beam~\cite{allen1992orbital,padgett1995poynting, allen_iv_1999}.
This is distinct from the spin angular momentum of a light field, which is associated with the polarisation state of the beam.
A prime example of light fields with orbital angular momentum are the cylindrically symmetric Laguerre-Gaussian (LG) transverse modes~\cite{allen1992orbital, allen_iv_1999}.
The transverse mode of a LG beam is described by two indices with the azimuthal, $l$, garnering the most interest~\cite{allen1992orbital}.
This index quantifies the amount of phase accumulation, of $2\pi l$, around the mode circumference~\cite{ allen1992orbital}.
Such fields have an orbital angular momentum of $l\,\hbar$ per photon.

Applications for LG modes have been found in many areas, including optical manipulation of atoms and mesoscopic particles~\cite{padgett2011tweezers}, where they lead to rotation of trapped structures, optical communication~\cite{willner2015optical, wang2012terabit}, where they can enhance data transmission security and quantum communication~\cite{mair2001entanglement, sit2017high}, where they offer access to a higher dimensional Hilbert Space enabling more efficient use of communication channels. 
Furthermore, the  LG mode with $l\,{=}\,1$ is central to super-resolution microscopy~\cite{willig2006sted} where the singularity at the centre of the beam is exploited to improve resolution to achieve nanometre level imaging\cite{scheiderer2025minflux,lukinavivcius2024stimulated}.

Though most of the interest has focused on LG beams with integer values of $l$, there does not need to be a multiple of $2\pi$ around the mode circumference. These light fields may possess a non-integer, or fractional, azimuthal index. 
Such fields may be decomposed into combinations of LG modes with integer azimuthal indices of varying amplitude~\cite{zhang2021review}. 
Therefore, the azimuthal index is in fact a continuous value and may take on any value.

A precise measurement of the azimuthal index, $l$, is a pertinent requirement in a number of the aforementioned applications~\cite{gibson2004free, willner2015optical,cheng2025metrology}.
The determination of the azimuthal index of LG modes has been explored at both the classical and single photon level~\cite{galvez2011interferometric, ren2021polarization, wang2022spin, leach2002measuring}.
Typical methods for determining the azimuthal index of a beam include: diffraction from an aperture~\cite{sztul2006double, mourka2011visualization}, modified Mach–Zehnder interferometers~\cite{leach2002measuring, huang2012measuring, li2015measuring, xu2024topological}, and more recently, statistical analysis or machine learning~\cite{liu2019superhigh, jing2021recognizing}.
For classical beams, the majority of these methods focus on determining integer $l$ and typically find it challenging to measure fractional $l$.
The most precise measurements to date for measuring fractional azimuthal index $l$ may resolve changes to an accuracy of 0.1 using a neural network~\cite{jing2021recognizing}, and 0.05 when using interference fringe dislocation~\cite{Shikder2023}.
Measurement beyond this precision remains a highly desirable, but unresolved, goal.

Here, we focus on the precision measurement of $l$, which is based on light propagation through disordered media resulting in speckle patterns.
Speckle patterns are generally regarded as a randomisation process of the optical field and have been viewed as deleterious to optical measurements, destroying information contained within the light field.
However, these interference patterns are deterministic and rich in information about the source of light, disordered medium, and environment~\cite{metzger2017harnessing, bruce2019overcoming, trivedi20193d, tran2020utilizing, barrios2024highly, facchin2023measuring}.
While analysis of light fields possessing OAM has already been investigated~\cite{Mazilu2012, bianchetti2019}, it has mainly focused on demonstrating classification of light fields with integer azimuthal indices~\cite{Fickler2017, Gong2019, Zhang2022, Ma2024}.
Here, we advance the precision of azimuthal index measurements using this innovative approach of speckle metrology, measuring non-integer values of the azimuthal index to an unprecedented measurement precision of $l$ of $2\,{\times}\,10^{-5}$, a three-order of magnitude improvement in comparison to other methods. 
In itself, this significant advance may allow for a higher density and increased data transfer rates in optical communications, and enhanced manipulation and control of the motion of trapped mesoscopic particles.

In this study, we exploit this precision measurement of azimuthal index to realise a novel approach to measure refractive index.
Our method of LG beam generation involves mode conversion with a microscopic spiral phase plate (SPP) immersed in the medium of interest within a microfluidic channel. Our approach samples a volume of 300\,pL. 
The mode conversion is dependent upon the refractive index of both the plate and the surrounding environment. 
Thus, we leverage our speckle based measurement of $l$ to, in turn, determine the refractive index of a liquid sample surrounding the SPP, with a best measured precision of $6.4\,{\times}\,10^{-7}$\,Refractive Index Units (RIU), which is at the shot-noise limit of the system.
We note that schemes have been used for refractive index sensing in bulk media, using the interference patterns of an LG beam with a co-aligned reference beam~\cite{dorrah2018experimental, meglinski2024phase, wu2024detecting}. 
Precisions of $5\,{\times}\,10^{-6}$\,RIU~\cite{meglinski2024phase} and $3.46\,{\times}\,10^{-7}$\,RIU~\cite{wu2024detecting} have been reported.
In contrast to these studies, our approach is based on a direct measurement of the LG beam, obviating the need for any reference field for interference.
To minimise the volume of sample required for testing, we utilise a microfluidic environment with 50\,$\mu$m diameter SPPs allowing refractive index sensing of picolitre sample volumes.
The direct nature of our measurement has multiple advantages: it obviates the requirement of a reference beam and thus alignment of an interferometer, interrogates picolitre volumes (rather than a bulk measurement), and may be readily multiplexed within the microfluidic chamber.



\section*{Principle of Operation}   \label{sec:design}
\begin{figure}[t]
    \centering
    \includegraphics[width=13cm]{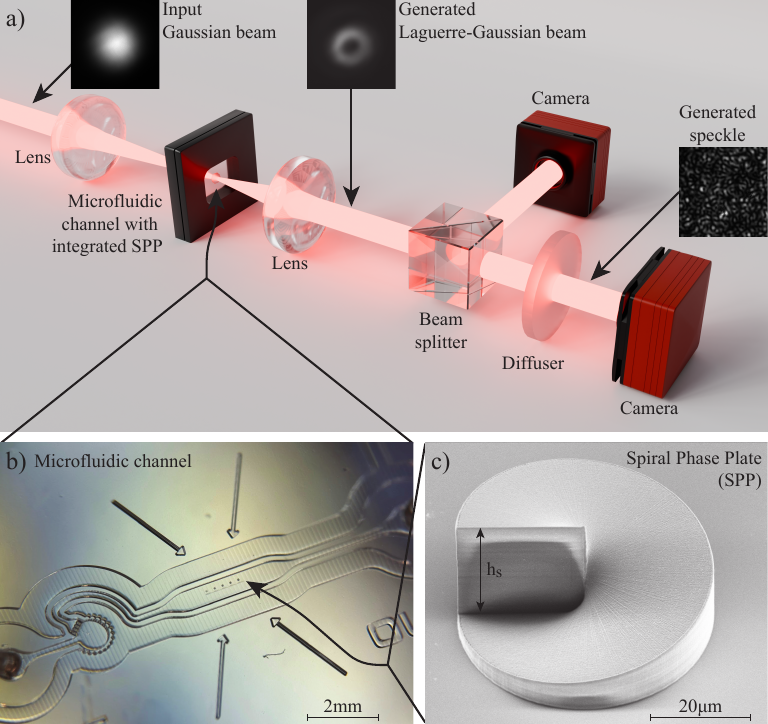}
    \caption{a) Experimental diagram showing a Gaussian beam passing through an SPP, transforming it into a Laguerre-Gaussian (LG) beam. 
    The presence of liquid samples of differing refractive index around the SPP changes the resultant azimuthal index of the beam.
    To accurately measure the azimuthal index of the resultant beam, a speckle pattern is generated using a diffuser and imaged on a camera. 
    The intensity profile of the LG beam prior to the diffuser is also imaged using a camera for reference.
    b) Shows a microscope image of the microfluidic chip with the SPPs at the centre.
    c) Shows a scanning electron microscope image of one of the SPPs and its dimensions. 
    The height of the phase plate step is denoted $h_s$. 
    }
    \label{fig:ExpDiagram}
\end{figure}

At the heart of our sensing approach is mode conversion of a Gaussian beam to a Laguerre-Gaussian (LG) beam and subsequent determination of the azimuthal phase accrued in the process.
We achieve this by imparting an azimuthally varying wavefront phase retardation upon an input Gaussian beam~\cite{beijersbergen1994helical, cano2018dynamic, de2023generation}, using a microfabricated spiral phase plate (SPP), shown in Fig.~\ref{fig:ExpDiagram}\,a) and b), a scanning electron microscope image of which is shown in Fig.~\ref{fig:ExpDiagram}\,c).
These SPPs create LG beams with an azimuthal index, $l$, and a radial index of $p\,{=}\,0$, see Fig.~\ref{fig:ExpDiagram}\,a) (inset).
 
After passing through the SPP, the azimuthal phase, $\phi$, accumulated around the beam is given by the difference in optical path length accrued around the wavefront, given by $\phi\,{=}\,2\pi h_s(n_\textrm{SPP}-n)/\lambda_0$ for a laser with a vacuum wavelength of $\lambda_0$ passing through a SPP of refractive index $n_\textrm{SPP}$ with a step height $h_s$ (see Fig.~\ref{fig:ExpDiagram}\,c), surrounded by a sample liquid of refractive index $n$.
Thus, the azimuthal index of the generated beam is:
\begin{equation}     \label{eqn:l}
    l = \frac{\phi}{2\pi} = \frac{h_s(n_\textrm{SPP} - n)}{\lambda_0}.
\end{equation}
Precise measurement of the azimuthal index, $l$, of the generated LG beam allows calculation of the refractive index of the sample via:
\begin{equation}    \label{eqn:n}
   n =  n_\textrm{SPP} - \frac{l \lambda}{h_s},
\end{equation}
allowing us to measure the refractive index of the sample media surrounding the SPP.

Figure~\ref{fig:ExpDiagram}\,a) shows the experimental diagram in which a laser with an input Gaussian spatial profile passes through an SPP, transforming it into a LG beam. 
The refractive index of the sample surrounding the SPP, $n$, determines the azimuthal index of the generated LG beam.
We use speckle metrology to precisely measure the azimuthal index of the beam, whereby a speckle pattern is generated via a diffuser then imaged on a camera.
A diffuser is used as it scatters the beam once, which is sufficient to extract maximum information regarding the spatial mode of a beam~\cite{facchin2024determining}.
This is in contrast to path-dependent measurements, such as wavelength or displacement, that gain sensitivity from multiple scattering events, utilising multimode fibre or integrating spheres to produce speckle patterns.
The benefit of utilising a single scatterer, such as a diffuser, is that it minimises the speckle measurement sensitivity to these other quantities, wavelength or displacement.
Principal component analysis (PCA) is utilised to extract the azimuthal index of the generated beam, and refractive index of the sample, the method of which is discussed in the methods section, and theoretical analysis of the process presented in the supplementary material, Sec.~\textcolor{blue}{1}.
In summary, a set of speckle images is required at known azimuthal/refractive indices to train the principal component analysis, after which the azimuthal index of an unknown LG beam, or the refractive index of an unknown sample, can then be measured. It is not necessary to measure every azimuthal index value. Rather, we may interpolate between training points in PCA space to determine the value of $l$~\cite{bruce2019overcoming}. 
The noise limit of our measurement is discussed in detail in the supplementary information, Sec.~\textcolor{blue}{4}.


\section*{Results}
\subsection*{Measurement Precision}
We test the precision of our method at two different wavelengths of 1064\,nm and 532\,nm, with two different SPPs with heights of $h_s\,{=}\,10.34\,\mu$m and $h_s\,{=}\,25.85\,\mu$m, designed to generate Laguerre-Gaussian (LG) beams with azimuthal index of $l\,{=}\,2$ and $l\,{=}\,5$ respectively in heavy-water D\textsubscript{2}O at a wavelength of 1064\,nm (refractive index $n\,{=}\,1.3211$ ~\cite{kedenburg2012linear}).
These SPPs are referred to as SPP2 and SPP5, respectively.

The following procedure was used to determine the measurement precision of both the azimuthal and refractive index.
For each SPP, the system was trained for refractive indices between 1.3328 and 1.3403 for which sucrose samples were used (0\,\% - 5\,\% sucrose solution), although any refractive index range can be chosen.
Following training, precision measurements were taken to ascertain the azimuthal and refractive index measurement precision.
Eight solutions within the training refractive index range were chosen, four of which were at the same refractive index as the training set, and four that were not.
For each of these eight solutions, 10 measurements of azimuthal and refractive index were made, the standard deviation of which gives an indication of the precision of each measurements.
This process, of training followed by eight precision measurements, is repeated 4-5 times, giving roughly 32-40 precision measurements for a single plate.
Averaging over the 32-40 precision measurements gives an average precision for the azimuthal and refractive index measurements for each plate.
The averaged precision is presented with errors, which are the standard error in the mean of these measurements.

\begin{figure}[t]
    \centering
    \includegraphics[width=8.0cm]{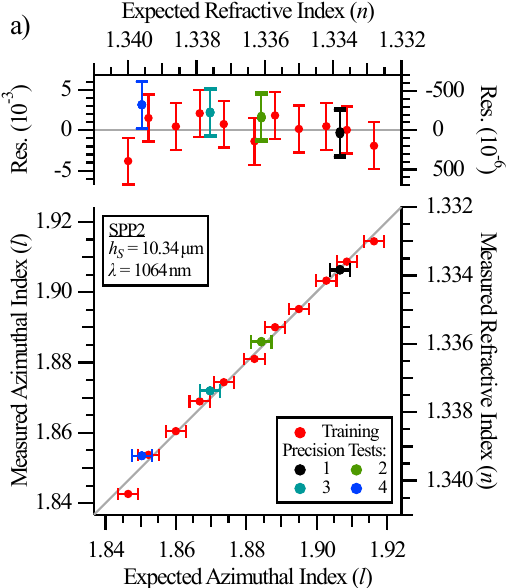}
    \hspace{1cm}
    \includegraphics[width=8.0cm]{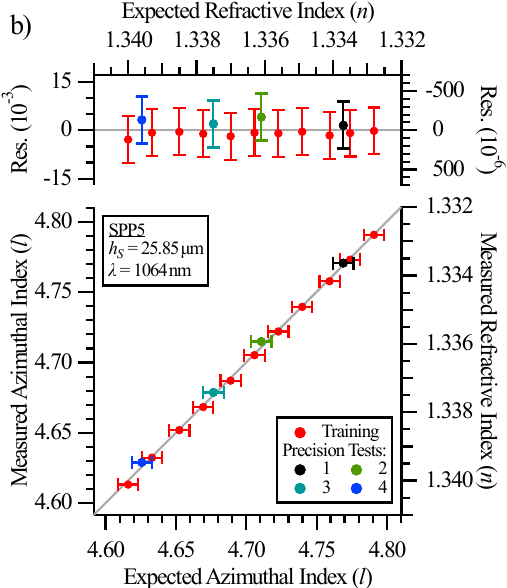}
    \caption{
    Measured azimuthal index as a function of the expected azimuthal index as well as the extracted refractive index measurements.
    The upper panel shows the precision of the measurements by presenting residuals between the measurements and the expected values. 
    a) Measurements for SPP2 with $h_s\,{=}\,10.34\,\mu$m.
    b) Measurements for SPP5 with $h_s\,{=}\,25.85\,\mu$m. 
    }
    \label{fig:precision}
\end{figure}

Figures~\ref{fig:precision}\,a) and b) shows the training and azimuthal and refractive index precision measurements for SPP2 and SPP5, respectively.
The training data fall around the line of expectation (grey), with the precision tests clumping tightly (10 measurements at each refractive index).
The error bars are based upon the measurement accuracy of the commercial refractometer (${\pm}\,3\,{\times}\,10^{-4}$\,RIU) used to determine the refractive index of the training solution (MyBrix, Mettler Toledo).
To better illustrate the excellent measurement precision, residuals are displayed showing the difference between the measurements and the expected values, from which minimal deviation from the expected values is observed.

At an input wavelength of 1064\,nm, for SPP2 an average refractive index precision of $\left(7.4\,{\pm}\,0.5\right)\,{\times}\,10^{-6}$\,RIU was achieved, from which SPP5 improved upon achieving an average precision of $\left(3.4\,{\pm}\,0.4\right)\,{\times}\,10^{-6}$\,RIU.
The best precisions measured for each plate was $3.3\,{\times}\,10^{-6}$ and $0.7\,{\times}\,10^{-6}$\,RIU for SPP2 and SPP5 respectively.
The precision of the system is expected to increase linearly with plate height, $h_s$ as described by Eqn.~\ref{eqn:l}, and discussed in more detail in the supplementary information, Sec.~\textcolor{blue}{2.1}. 
Thus, we expect an increase in the precision of SPP5 compared to SPP2 of 2.5-fold based on their respective heights of $h_s\,{=}\,25.85\,\mu$m and $h_s\,{=}\,10.34\,\mu$m.
Our measured precisions give an average ratio of $2.2\,{\pm}\,0.3$ between the two plates, which agrees well with expectation. 
The refractive index precision is limited by shot noise of the detected light, discussed in detail in supplementary information, Sec.~\textcolor{blue}{4.3}.
 
At 1064\,nm for both SPP2 and SPP5, the average precision of the azimuthal index measurements respectively is $\left(7.2\,{\pm}\,0.5\right)\,{\times}\,10^{-5}$ and $\left(8.6\,{\pm}\,1.0\right)\,{\times}\,10^{-5}$. 
The best precision of the azimuthal index measurements is $3.2\,{\times}\,10^{-5}$ and $1.8\,{\times}\,10^{-5}$ for the SPP2 and SPP5 respectively.
With increasing azimuthal index, $l$, the grain size of the speckle pattern reduces~\cite{hu2020does,li2015study}.
In path-dependent speckle measurements, such as wavelength or displacement measurements, a finer grain size typically leads to higher measurement sensitivity~\cite{facchin2024determining}.
However, for azimuthal index measurements, the sensitivity of the speckle measurement to changes in azimuthal index is independent of speckle grain size, thus the magnitude of the azimuthal index~\cite{facchin2024determining} remains constant, further details can be found in the supplementary information, Sec~\textcolor{blue}{1}. 
This is observed in the ratio of the measured azimuthal index precisions between SPP2 and SPP5 to be $0.9\,{\pm}\,0.1$, in line with the expected ratio of $1$.

\begin{figure}[t]
    \centering
    \includegraphics[width=8.0cm]{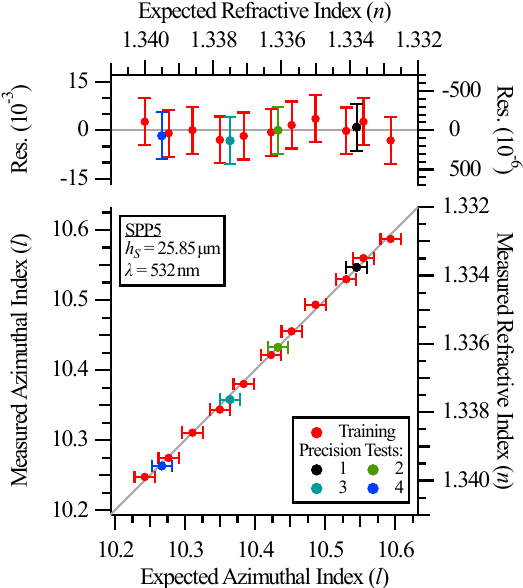}
    \caption{
    Measurements at 532\,nm showing measured azimuthal index as a function of the expected azimuthal index as well as the extracted refractive index.
    The upper panel shows the precision of the measurements by presenting residuals between the measurements and the expected values. 
    }
    \label{fig:Precision532nm}
\end{figure}

The average refractive index precision is further improved by moving to a shorter input wavelength, leading to a larger phase shift being accrued around the beam circumference after passing through the SPP, leading to a generated beam with a higher azimuthal index.
This is seen by the $1/\lambda$ proportionality in Eqn.~\ref{eqn:l}, and discussed in more detail in the supplementary information, Sec.~\textcolor{blue}{2.2}.
Figure~\ref{fig:Precision532nm} shows the training and precision measurements for SPP5 at the input wavelength of 532\,nm.
For SPP5, at 532\,nm, an average precision of $\left(2.2\,{\pm}\,0.3\right)\,{\times}\,10^{-6}$\,RIU was achieved, with the best precision observed being $0.64\,{\times}\,10^{-6}$.
When compared to the 1064\,nm SPP5 results, we expect a two-fold improvement in precision using 532\,nm due to the  $1/\lambda$ proportionality in Eqn.~\ref{eqn:l}. This is discussed in more detail in the supplementary information, Sec.~\textcolor{blue}{2.2}.
Comparing measured precisions for SPP5 at 532\,nm and 1064\,nm, we observe a ratio of $1.6\,{\pm}\,0.3$ which agrees well with expectation.

At 532\,nm, for SPP5, the average precision of the azimuthal index is $\left(10.6\,{\pm}\,1.7\right)\,{\times}\,10^{-5}$, with the best measurements being $3.1\,{\times}\,10^{-5}$.
To our knowledge, these measurements for fractional azimuthal index  are the most precise to date.
Similar to the expected azimuthal index precision ratio between SPP2 and SPP5 at 1064\,nm discussed above, we also expect an azimuthal index precision ratio of $1$ between SPP5 at 1064\,nm and SPP5 at 532\,nm, for which we measured a ratio of $0.8\,{\pm}\,0.2$ in line with these expectations.

\subsection*{Precise refractive index measurements of haemoglobin}
To validate the applicability of the system, we performed precision measurements on ultra-low volumes of a biological sample, namely haemoglobin.
The volume of the sample needed to fill the microfluidic chamber containing multiple SPPs, seen in Fig.~\ref{fig:ExpDiagram} b), is only 22\,$\mu$L.
However, the sample volume that the light traverses through is only 300\,pL, given by the chamber height of 160\,$\mu$m and 50\,$\mu$m diameter of a SPP. 
In future, the chamber size could be reduced considerably to accommodate a single SPP, reducing the sample volume towards the nanolitre or picolitre regime.

\begin{figure}[t]
    \centering
    \includegraphics[width=8.0cm]{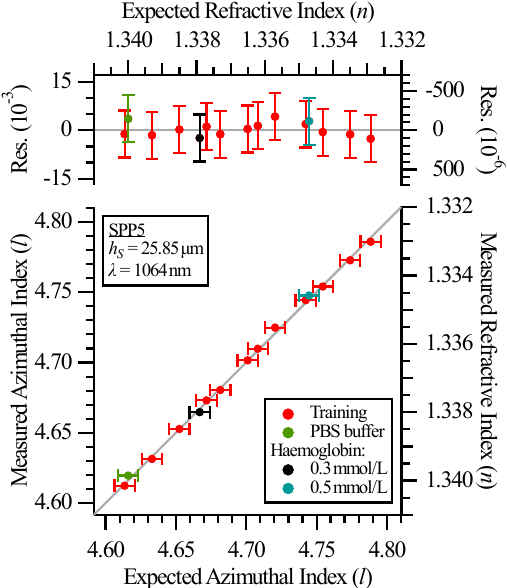}
    \caption{
    Study of biological measurements based on haemoglobin are shown in Black, Green, Cyan using SPP5 at 1064\,nm.
    Measured azimuthal index as a function of the expected azimuthal index as well as the extracted refractive index measurements.
    The upper panel shows the precision of the measurements by presenting residuals between the measurements and the expected values. 
    }
    \label{fig:biodata}
\end{figure}

In a biological context, precise refractive index measurements have been applied to protein characterisation studies by determining the protein refractive index increment~\cite{zhao2011distribution}.
The precise measurement of protein refractive index has significant applications as a non-destructive form of protein analysis in a range of industries, such as food and beverage as well as pharmaceutical drug testing~\cite{seki2016refractive, mohan2019refractive}. 
Recently, the refractive index increment of low concentrations of protein was measured using precision interferometry to study their conformational state~\cite{matveyeva2020precision}. 
It is worth noting the use of interferometry to measure refractive index differs from our described methods.  

To test our system in a biological setting, we first trained the system using sucrose solution, then measured the refractive index of three biological samples.
Figure~\ref{fig:biodata} shows results using SPP5 at 1064\,nm. 
We note that similar results were obtained using SPP2 at 1064\,nm.
In our case, these samples were two different concentrations of haemoglobin (0.3\,mmol/L and 0.5\,mmol/L) which were dissolved in a PBS (phosphate buffered saline) buffer, with a blank PBS buffer being the third solution.
The refractive index measured for the biological sample was within 0.02\,\% of the actual value, as measured by the commercial refractometer, as seen in Fig.~\ref{fig:biodata}.
This demonstrates that by using our speckle-based measurements of azimuthal index, we could accurately determine the refractive index of any unknown biological sample to a very high precision.


\section*{Discussion}
We have demonstrated ultra-precise measurement of the fractional azimuthal index of a Laguerre-Gaussian (LG) beam to a precision of $2\,{\times}\,10^{-5}$ using speckle metrology.
This precision exceeds the previous best azimuthal index measurement by over three orders of magnitude, and will significantly impact fields that utilise OAM, such as optical communication. 
Furthermore, in applications that utilise optical manipulation of atoms and mesoscopic particles, this may lead to finer control of trapped particles/structures.

As part of our study, we demonstrated the ability to create LG beams with finely controlled azimuthal index, with minimum step size of 0.002.
Here the concentration of the sucrose solution surrounding the SPP determined the azimuthal index of the generated beam.
Typically, a spatial light modulator (SLM) is used to create beams with variable azimuthal index.
However, we remark that the minimum azimuthal index step size demonstrated here of 0.002 requires a larger bit depth than that of a standard SLM (8-bit).
Furthermore, to consolidate and verify the azimuthal index measurement precision of $2\,{\times}\,10^{-5}$ demonstrated here, a 16-bit SLM is required to achieve an azimuthal index step size at this level.
That is, the azimuthal index measurement precision demonstrated here is already at the limit of what is possible using the highest bit-depth SLM commercially available, a 16-bit SLM (for example, Meadowlark Optics, HSPDM512 1064-PCIe 16-bit).
SLMs with such a high bit-depth are not common and thus pose a challenge to test with this approach.
We note direct-write lithography for microfabrication with such a high bit-depth may be an alternative approach for fine control of holograms of specific azimuthal index (e.g. the DWL 66+ Laser Lithography System has a grayscale capability of 65,536 levels).

We utilise this determination of azimuthal index to measure the refractive index of the sucrose solution surrounding the SPP, achieving a best precision of $6.4\,{\times}\,10^{-7}$\,RIU.
This compares favourably to existing commercial and research technologies.
Commercial refractometers typically utilise total internal reflection, achieving precisions of $10^{-5}$\,RIU.
Comparing our approach more broadly, plasmonic refractive index sensors~\cite{white2008performance, xu2019optical} reach precisions of $10^{-6}$\,RIU, but often have limited measurement range due to properties of the plasmonic structure.
Whispering gallery optical resonators have achieved a precision of $2.8\,{\times}\,10^{-7}$\,RIU~\cite{li2010characterization}, while polarization-dependent absorption of graphene has achieved $1.7\,{\times}\,10^{-8}$\,RIU precision enabling single cell detection~\cite{xing2014ultrasensitive}.
The precision of our refractive index measurements can be improved by increasing the azimuthal index of the generated beam via two routes: utilising shorter wavelengths and using SPPs with greater heights.
Moving to a laser wavelength of 385\,nm and a SPP height of 258.5\,$\mu$m would lead to a 30-fold improvement to the refractive index measurements compared to our 1064\,nm measurement with SPP5.
This would enable a measurement precision of $3\,{\times}\,10^{-8}$.
See the supplementary information, Sec.~\textcolor{blue}{2.1} \& \textcolor{blue}{2.2} for a more detailed discussion.

To demonstrate the potential of our approach for biological studies, we measured the refractive index of haemoglobin. Haemoglobin, the oxygen-binding protein found within red blood cells, is responsible for transporting oxygen throughout the body, and its optical properties, including refractive index, are of diagnostic relevance~\cite{serebrennikova2015characterization}. Specifically, the refractive index of haemogloin is noted as an indicator of health and is presently a challenging parameter to measure accurately~\cite{doi:10.1021/acsomega.2c00746}. Our study showed agreement of the refractive index of haemoglobin to within 0.02\,\% of that measured with a commercial refractometer, which also agrees with literature values~\cite{zhernovaya2011refractive}. Additionally, we recorded the refractive index of haemoglobin across a range of concentrations, with results consistent with previously reported values~\cite{jin2006refractive}.  
Our demonstrated capability to perform these measurements with high accuracy in such miniscule sample volumes, offers a valuable approach for precise optical characterisation that may be extended to other liquid or gaseous analytes in future studies. Specifically, measurements of refractive index are not only of importance for biology but may also serve as a quality control parameter in industries such as food and beverage production (e.g. sugar content of drinks), pharmacology, or chemical analysis~\cite{Jaywant2022, xu2019optical}.

The microfluidic environment within which the refractive index measurements were undertaken is worthy of discussion. 
We used a chamber with a sample volume of 22\,$\mu$L, within which the SPP and laser sampled a dramatically smaller volume of approximately 300\,pL.
We expect that the total sample volume could be reduced to 50\,pL by designing the the chamber to match the volume sampled by the SPP and laser.
Interestingly, metasurfaces, such as sub-wavelength nano-pillar quadrumer, may offer a route for a further reduction in sample volume to hundreds of attolitres~\cite{chen2024highly}. 
Turning to miniaturisation of the optical footprint, improved stability could be attained by utilising an SPP  directly  printed onto the distal end of an optical fibre~\cite{weber2017single}.
Importantly, our scheme could be used to overcome challenges associated with biological samples that are found in extremely low quantities, allowing for ultra-low volume measurements of low volume nucleic acids~\cite{desjardins2010nanodrop} or low molecular weight compounds~\cite{DAVIS2000348}.
The application of refractive index measurements in single-cell protein analysis~\cite{zhang2017quantitative} suggests a promising avenue for further study in these areas.

To increase throughput and enable detection of spatial variations in refractive index, multiple measurements could be made on a single chip.
This could be achieved either with arrays of SPPs in parallel microfluidic channels to measure multiple samples simultaneously, or by  integrating multiple SPPs within a single chamber to measure spatial variations in refractive index within a sample.
Each SPP could be interrogated either sequentially, or simultaneously by assigning distinct interrogation wavelengths to each SPP and separating the outputs with wavelength filtering after the sample, for example using an optical frequency comb as the light source.
Furthermore, these refractive index measurements can be taken at high speeds, only limited by the exposure time of the camera to ensure the speckle pattern is captured with sufficient SNR, enabling real-time measurements.
For the demonstration here, the exposure time used was 20\,ms leading to a 50\,Hz measurement rate, which could be improved with a higher quality camera or higher optical power. 
In a microfluidic environment, refractive index gradients can be highly informative.
For example, spatio-temporal differences in refractive index occur during early embryo development and measurement of other such gradients critical in immune cell response, and development of metastatic cancer would be interesting to investigate~\cite{keenan2008biomolecular, Dwapanyin:23}.
Additionally, refractive index gradients in microfluidic environments can identify molecular transport (mixing) between adjacent laminar flows and allow investigation of molecular diffusion at flow boundaries~\cite{RevModPhys.77.977, sung2014three, costin2002measuring}.
Integrating real-time refractive index measurements into a multiplexed system would allow monitoring of refractive index gradients as a biological system evolves.

In summary, we have demonstrated refractive index measurements of unprecedented precision in ultra-low sample volumes, enabled by speckle-based determination of the fractional azimuthal index. This approach outperforms existing commercial systems and is highly competitive with other research-grade refractometers but also offers a scalable platform for multiplexed, high-speed measurements in a microfluidic environment. The ability to resolve refractive index gradients in real time opens new opportunities for studying dynamic biological processes, such as molecular diffusion and protein signalling, as well as for monitoring chemical and industrial systems where quality control is critical. By combining very-high sensitivity with the potential for miniaturisation, high throughput, and spatio-temporal resolution, our methodology establishes a new application by utilising precision measurement of light fields with orbital angular momentum. This offers a foundation for next-generation refractive index sensing across biology, chemistry, and industry.



\section*{Methods}

\subsection*{Spiral Phase Plate (SPP) fabrication}  \label{sec:spp_fab}
Computer-aided design software (CAD), Solidworks (Dassault Systèmes), was used to generate parametric three-dimensional model geometries, and exported to STL format.
STL files were imported into the computer-aided manufacturing (CAM) software, DeScribe (Nanoscribe GmbH).
Layer heights were set between 10\,nm and 50\,nm depending on plate height, optimised within DeScribe to ensure smooth surfaces on the spiral ramp, with the maximum edge length of any horizontal layer plane kept below 200\,nm.
The exposure parameters were individually optimised for each layer height.
The devices were printed as solid bodies using the Nanoscribe Photonic Professional GT+ operating in dip-in mode with IP-DIP resin deposited onto a 25\,mm square fused silica slide.
The printer was equipped with a 63$\times$ objective (all Nanoscribe GmbH).
Following exposure, devices were developed in approximately 20\,mL of SU-8 developer (Kayaku Advanced Materials Inc.) for approximately 6\,minutes, and blow dried with nitrogen.
The resulting SPP were imaged with a scanning electron microscope, which is shown in Fig.~\ref{fig:ExpDiagram}\,c). 
The fabrication accuracy is discussed in the supplementary information, Sec.~\textcolor{blue}{3.4}.
The refractive index of the SPP polymer at 1064\,nm is $n_\textrm{SPP}\,{=}\,1.53$~\cite{gissibl2017refractive, dottermusch2019exposure}, and at the wavelength of 532\,nm the refractive index is $n_\textrm{SPP}\,{=}\,1.55$~\cite{gissibl2017refractive, dottermusch2019exposure}.
The refractive index wavelength dependence~\cite{schmid2019optical}, along with thermal expansion properties~\cite{qu2017micro}, of IP-DIP is discussed in more detail in the supplementary information, Sec.~\textcolor{blue}{3.2} and \textcolor{blue}{3.5} respectively.

\subsection*{Fabrication of the microfluidic chamber}   \label{sec:pdms_fab} 
The microfluidic system is based on polydimethylsiloxane (PDMS) polymer (SYLGARD™ 184, Dow Corning), fabricated using standard photolithography techniques~\cite{mcdonald2002poly}.
Channel structures are patterned into the underside of a PDMS slab through casting onto a template. 
The negative template consisted of a raised SU-8 3050 photoresist (Kayaku Advanced Materials Inc.) structures patterned on a flat Silicon wafer substrate. SU-8 3050 was spin-coated onto a 4-inch silicon wafer to a thickness of 160\,$\mu$m, the SU-8 was then soft baked to remove solvents before patterning using a maskless aligner (MLA 150, Heidelberg instruments).
Crosslinking of patterned resist was initiated with a post-exposure bake, followed by removal of unexposed resist through sonication in SU-8 Developer. 
The wafer was hard baked to increase mechanical stability and remove residual solvents, then silanized to aid with mould release, through exposure to trichloro(1,1,2,2-perfluoocytl)silane (Sigma-Aldrich) vapour in a desiccator, resulting in a robust, re-usable mould for PDMS casting.

PDMS prepolymer was mixed in a Planetary Centrifugal Vacuum Mixer (THINKY ARV-310) with curing agent in a 1:10 ratio by weight. 
PDMS was cast on the master template and cured at 130\,$^\circ$C in a convection oven for 25\,min. 
The cured PDMS was peeled off the mould, cut to size, and punched with 0.75\,mm interface holes. 
The system consisted of a single fluid channel, with an expanded chamber that is placed over a set of SPPs, see the schematic in Fig.~\ref{fig:ExpDiagram}\,b). 
A bubble trap ~\cite{lochovsky2012bubbles} was included prior to the phase plate chamber to reduce the impact of bubbles that may otherwise cause unwanted scattering of the beam. The fluidic system included an integrated vacuum chuck (offset channels) facilitating bubble absorption and easy reuse through substrate changeover, with negative pressure applied through an attached syringe housed in a custom-built jig. 
The PDMS chip was aligned with SPPs using a dissecting microscope, and flow was achieved by connecting a negative pressure pump to the channel outlet. The flow rate in the microfluidics channels was set at 2\,cc/min, this rate can be lowered or increased by changing the pressure in the negative pressure pump.

\begin{figure}[t]
    \centering
    \includegraphics[width=13cm]{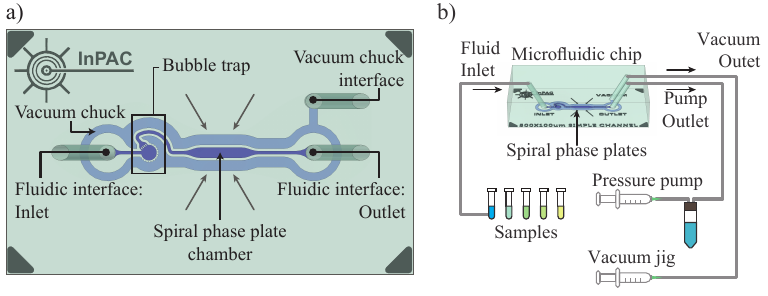}
    \caption{
    a) Schematic of the microfluidic chip with marked components. 
    A microscope image of this chip is shown in Fig.~\ref{fig:ExpDiagram}\,b).
    The fluids enter from the inlet and travel through the purple-coloured region that houses the SPPs and later exit through the outlet. 
    The blue-coloured region is connected to the vacuum jig designed to encourage reversible sealing of the PDMS to glass.
    b) A single inlet is connected to sample Eppendorf’s containing liquid samples of different refractive index through Tygon tubing. 
    The fluid outlet is connected to a basic negative pressure pump consisting of a syringe and 50\,mL Falcon tube.
    Vacuum is supplied by a drawn 50\,mL syringe held open by a jig.
    Withdrawing air from the pressure pump creates a pressure differential driving flow. 
    Withdrawal of 1\,mL of air is sufficient to maintain flow rates of approximately 50\,$\mu$L/min for the duration of the experiment, while fluid switching is achieved by clamping the inlet during transfer between sample tubes.
    }
    \label{fig:MicroSchem}
\end{figure}

\subsection*{Sucrose solution preparation}  \label{sec:sucrose}  
Sucrose solutions of various refractive indices were prepared by mixing sucrose with water.
For example, a 5\,\% by mass sucrose solution (5\,\% by mass sucrose solution, known as 5\,Brix), 5\,g of sucrose (Chem-supply sucrose) was mixed with 95\,g of Mili-Q water (Merk Mili-Q-IQ 7000).
The solution was mixed on a magnetic stirrer (IKA, RCT basic).
The refractive index of each of the sucrose solutions made was checked using a handheld refractometer (MyBrix, Mettler Toledo).
These solutions were then stored in a 100\,mL capped glass bottle for repeated use. 
Prior to the experiment, the solutions were filtered with a 0.22\,$\mu$m syringe filter (Millex Millipore) into a 1.5\,mL Eppendorf tube.
The refractive index of the samples was remeasured every time at the start of the experiment to ensure any changes in the refractive index over time were taken into account.  
The sucrose samples refractive index dependence upon wavelength~\cite{ghosh1997sellmeier, ghosh1994temperature, bhar1980temperature, purwandari2021analyses} and temperatures~\cite{daimon2007measurement} is discussed in more detail in the supplementary information, Sec.~\textcolor{blue}{3.1} and \textcolor{blue}{3.3} respectively.

\subsection*{Haemoglobin samples preparation}  \label{sec:Hemo}
Ferrous haemoglobin (H0267, Sigma Aldrich) was dissolved in phosphate buffered saline (PBS) to concentrations of 0.3\,mmol/L and 0.5\,mmol/L.
Samples were briefly vortexed and filtered using a $0.22\,\mu$m filter (Millex Millipore).
A blank sample containing only PBS was also prepared. New samples were prepared prior to the start of each experiment, and the refractive index verified with a commercial refractometer (MyBrix, Mettler Toledo).
The haemoglobin samples refractive index dependence upon temperatures~\cite{lazareva2016temperature} is discussed in more detail in the supplementary information, Sec.~\textcolor{blue}{3.3}.

\subsection*{Optical measurements}  \label{sec:OptMeas}
The laser source used was IRCL-300-1064-S, CrystaLaser LC for measurements at 1064\,nm, which was focused onto the SPP using a plano-convex lens with a 100\,mm focal length (AC254-100-B-ML, Thorlabs), as seen in  Fig.~\ref{fig:ExpDiagram}\,a).
The SPP converts the input Gaussian spatial profile laser beam into a Laguerre-Gaussian (LG) beam shape. 
The transmission of such an SPP is greater than 95\%.
An aspheric Lens with a 11\,mm focal length (C220TMD-B, Thorlabs) was used to image the converted LG beam onto the camera imaging the beam. 
A charge-coupled device (CCD) camera (EO-1312M, Edmund Optics) and a complementary metal–oxide–semiconductor (CMOS)-camera (Alvium 1800 u-158m, Allied Vision) are used for the detection of the beam profile, as a reference image, and speckle patterns, respectively.
The effect of camera noise~\cite{Peterkovic2025} upon the measurement is discussed in the supplementary material, Sec.~\textcolor{blue}{4.4}.
The speckle pattern was generated using a disordered medium in front of the CMOS camera - in our case, this was a piece of scotch tape with $40\,\mu$m thickness.

For the measurements at input wavelength of 532\,nm a second harmonic generator (WH-0532-000-F-B-C) was used to convert a 1064\,nm laser source (Coherent Mephisto 2000) into one at 532\,nm. 
For a fair comparison, special care was taken to ensure that the beam size at the SPP was the same for both wavelengths; this was achieved by changing the beam size before the 100\,mm lens. 
It is also worth mentioning that at 532\,nm wavelength, the optics were switched to their counterparts with anti-reflection coating at 532\,nm.

The first step was to train the system in a desired range of refractive indicies, for this sucrose solutions of known refractive index were used. 
Training the system with 11 different sucrose solutions was found to be sufficient to reliably train the principal component analysis algorithm, as well as to keep the duration of the experiment small to minimise any drifting, which may result in calibration errors. 
These samples were introduced through the microfluidic environment, which housed the SPP within it. 
Since the SPP was immersed in a sample of interest with a particular refractive index, the produced LG beam depended on the refractive index and consequently, the speckle pattern depends on the refractive index of the sample as described by Eqn~\ref{eqn:n}.
Once the training in a particular region was achieved, the refractive index of any unknown solution in that region could be determined.

\subsection*{Determining the azimuthal index and refractive index using speckle}    \label{sec:Speckle}
The azimuthal index was determined based on the generated speckle patterns, which were analysed with the principal component analysis (PCA) technique. 
To train the PCA and determine the transformation matrix to extract the first principal component, a range of sucrose media with different refractive indices was passed through the microfluidic chip, immersing the SPPs.
This typically consisted of a image for 11 sucrose media of different refractive indices which were used to train the PCA to determine the transformation matrix and extract the first principal component of the data.
Typically, principal component 1 contained ${\approx}\,90$\,\% of the variance of the data, highlighting that the PCA is only dependent upon the change in azimuthal index and other environmental changes are not being relevant by this analysis.
The first principal component was compared to the known refractive index measured by using the commercial refractometer, with a fit to these two data sets taking the form of a linear relationship. 
With the trained transformation matrix, the samples of unknown refractive index values were measured, with their imaged speckle patterns input into the PCA and the refractive index produced.
Unknown refractive index samples were able to be measured via interpolation between training points due to the linearity of the PCA analysis and the known relationship between the first principal component and refractive index.
A detailed theoretical analysis of the PCA process presented in the supplementary material, Sec.~\textcolor{blue}{1}.

\bibliography{bibliography}

\section*{Acknowledgements}
K.D. acknowledges support from the Australian Research Council (FL210100099). 
This research was supported by the Australian Research Council Centre of Excellence in Optical Microcombs for Breakthrough Science (project number CE230100006) and funded by the Australian Government.
Device fabrication was conducted at the RMIT Micro Nano Research Facility (MNRF) in the Victorian Node of the Australian National Fabrication Facility (ANFF).
K.R.D. acknowledges support from National Health and Medical Research Council (APP2029067), K.R.D is supported by a Future Making Fellowship (University of Adelaide), and the Australian Research Council (FT240100291).
A.B. is supported by an ARC DECRA fellowship (DE230100964).
This work is supported by the US Office of Naval Research Global (N62909-25-1-2029).
We acknowledge Alexander Trowbridge for assistance in generating the experimental diagram.

\section*{Author contributions statement}
K.D. conceived and supervised the study.
A.A.P. and C.P. conducted the experiment(s), preliminary measurements by A.Z.
A.Z., C.S., A.B., and A.M. designed and fabricated the spiral phase plates. 
C.P. developed the speckle analysis approach and analysed the data. 
M.L. and K.R.D. developed the biological studies. 
K.D., C.P., A.A.P., and C.S. wrote the manuscript with contributions from other authors. 
All authors reviewed, edited, and approved the manuscript.

\section*{Additional information}
\textbf{Competing Interests:} The authors declare no competing interests.

\include{Supplementary_information}

\end{document}

%% file: Supplementary_information.tex
\section*{\centering Supplementary Information}
\section{Speckle Measurement Analysis} \label{sec:SpeckleAnalysis}
We consider the sensitivity of the speckle analysis using two different methods, which are discussed below.
The first method is based upon calculating the similarity of the speckle patterns, while the second method uses principal component analysis (PCA) to determine speckle sensitivity.

\subsection{Similarity comparison between LG Beams}
To determine the sensitivity of the speckle patterns for Laguerre-Gaussian beams with different azimuthal indices, we use the width of the similarity function as the figure of merit.
The similarity of the speckle patterns is also known as the Pearson correlation coefficient, or the zero-mean normalised correlation coefficient~\cite{facchin2024determining}.
For a speckle pattern that is fully developed~\cite{facchin2024determining}, the similarity, $S$, is defined as:
\begin{equation}    \label{eqn:similarity}
    S= \left| \frac{\langle \textbf{E}^\dagger \textbf{E}'  \rangle}{ \sqrt{\textbf{E}^\dagger \textbf{E}} \sqrt{\textbf{E}'^\dagger \textbf{E}'} } \right|^2,
\end{equation}
where $\textbf{E}$ and $\textbf{E}'$ are the electric fields of the speckle pattern, with $\textbf{E}$ representing the speckle pattern before a change is made, and $\textbf{E}'$ after the change is made, in this case the change would be a varying azimuthal index as a result of a change in refractive index.
$\dagger$ denotes the conjugate transpose of the electric field.
It can be shown that the similarity, $S$, is independent of the properties of the scatterer, thus, the electric fields defined in Eqn.~\ref{eqn:similarity} can be assumed to be 
the input electric fields before the scatterer.

For a Laguerre-Gaussian beam, the intensity normalised electric field is defined as~\cite{allen_iv_1999, berry_exact_2008, rosales-guzman_review_2018}:
\begin{equation} \label{eqn:LGElec}
    E(r,\phi, z) = \frac{C_{l,p}}{w(z)} \left(\frac{r \sqrt{2}}{w(z)}\right)^{|l|} \exp \left(-\left(\frac{r}{w(z)}\right)^2\right)  L_p^{|l|}\left[2 \left(\frac{r}{w(z)}\right)^2\right] \times \exp \left(i k \frac{r^2}{2 R(z)}\right) \exp \left(i l \phi\right) \exp (i \psi(z)),
\end{equation}
for a radial position $r$, azimuthal position $\phi$, and at $z$ position along the beam axis. Here $w(z)$ is the beam waist, $l$ is the azimuthal index of the Laguerre-Gaussian beam, $L_p^{|l|}\left[2 \left(\frac{r}{w(z)}\right)^2\right]$ are the generalised Laguerre polynomials, $k$ is the wave vector, $R(z)$ is the wavefront curvature, and $\psi(z)$ is the Gouy phase.
The normalisation factor, $C_{l,p}\,{=}\,\sqrt{\frac{2 p !}{\pi \left(p+\left|l\right|\right)!}}$, normalises the electric field such that $\int_0^{2\pi} \int_0^\infty r\left|E(r,\phi, z) \right| dr d\phi\,{=}\,1$.

For the experiments here, we are considering only Laguerre-Gaussian beams with $p\,{=}\,0$, thus $L_0^{|l|} \left(2 \left(\frac{r}{w_0}\right)^2\right)\,{=}\,1$.
We will also assume the beam is collimated, $R(z)\,{\rightarrow}\,\infty$, for simplicity.
In this limit, $w(z)\,{\rightarrow}\,w_0$, and $\psi(z)\,{\rightarrow}0$.
This is a valid assumption, as the Laguerre-Gaussian beam is collimated when incident upon the scattering media.
This assumption simplifies Eqn.~\ref{eqn:LGElec} to:
\begin{equation} \label{eqn:LGElecAss}
    E(r,\phi, z) = \frac{C_{l,0}}{w_0} \left(\frac{r \sqrt{2}}{w_0}\right)^{|l|} \exp \left(-\left(\frac{r}{w_0}\right)^2\right) \times \exp \left(i l \phi\right).
\end{equation}

\begin{figure}[t]
    \centering
    \includegraphics[width=6.5cm]{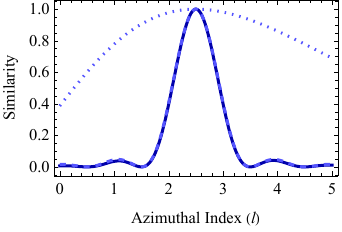}
    \hspace{1cm}
    \includegraphics[width=6.5cm]{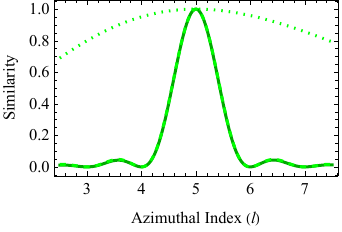}
    \caption{Theoretical similarity plots for reference beams of $l_0\,{=}\,2.5$, left, and $5.0$, right.
    The similarity is plotted from Eqn.~\ref{eqn:SimLG}, along with each of its main terms plotted separately.
    Solid line: Similarity given by Eqn.~\ref{eqn:SimLG}.
    Dashed line: First term of Eqn.~\ref{eqn:SimLG}, $\textrm{sinc}\left[ \pi \left(l - l_0\right) \right]^2.$
    Dotted line: Second term of Eqn.~\ref{eqn:SimLG}, ${\Gamma\left[1 + \frac{\left|l\right| - \left|l_0\right|}{2} \right]}/{\left(\Gamma\left[1 + \left|l\right| \right]  \,.\, \Gamma\left[1 + \left|l_0\right| \right]\right)}.$
    This shows the dominance of the $\textrm{sinc}^2$ term compared to the Gamma function factors.
    }
    \label{fig:TheorySim}
\end{figure}

From here, the similarity between a reference speckle pattern and a test speckle pattern with azimuthal indices of $l_0$ and $l$ respectively can be calculated via Eqn.~\ref{eqn:similarity} and substituting in Eqn.~\ref{eqn:LGElecAss}.
This gives the following expression:
\begin{equation}    \label{eqn:SimLG}
    S = \textrm{sinc}\left[ \pi \left(l - l_0\right) \right]^2 \times \frac{\Gamma\left[1 + \frac{\left|l\right| - \left|l_0\right|}{2} \right]}{\Gamma\left[1 + \left|l\right| \right]  \,.\, \Gamma\left[1 + \left|l_0\right| \right]}.
\end{equation}
The functional form of this is approximately a $\textrm{sinc}^2$ function with a broad envelope governed by the second term of the equation.
This can be seen in Fig.~\ref{fig:TheorySim}, which shows the dominance of the $\textrm{sinc}^2$ term compared to the Gamma function factors.
Under the approximation that the broad envelope is given by the Gamma functions is constant for $\left|l - l_0\right|\,{<}\,1$, the similarity can be approximated as:
\begin{equation}    \label{eqn:SimLGApprox}
    S \approx \textrm{sinc}\left[ \pi \left(l - l_0\right) \right]^2.
\end{equation}

There are a number of important aspects to note from Eqn.~\ref{eqn:SimLGApprox}.
The first is that this similarity expression is in agreement with that derived for a perfect vortex as derived theoretically and shown experimentally in Refs.~\cite{bianchetti2019, facchin2024determining}.
Second, Laguerre-Gaussian beams, and thus their speckle patterns, that differ by an azimuthal index of 1 are uncorrelated with each other.
This is evident by the fact that $S\,{=}\,0$ for $\left|l - l_0\right|\,{=}\,m$ where $m\,{\in}\,\mathbb{Z}$.
Third, is that the width of the similarity curve is independent of the reference azimuthal index value $l_0$, with a constant full width at half maximum of 0.886 in units of azimuthal index.
Phrasing this a different way, the rate of decoherence between speckle patterns with changing azimuthal index does not depend upon the azimuthal index. 
Thus, the sensitivity of speckle patterns to changes in the azimuthal index is independent of the absolute value of the azimuthal index of the beam.

\subsection{Principal component analysis (PCA) comparison between LG Beams}
While analysis of the similarity of the speckle patterns gives us insight to the rate at which speckle patterns decorrelate with changing azimuthal index, and hence sensitivity to changing azimuthal index.
However for ease of application in this work, we utilised principal component analysis (PCA) for analysing the speckle patterns.
As such, here we seek to understand the output dependency of PCA upon input speckle patterns generated from beams with differing azimuthal indices.
We do this numerically with the process outlined below.

The first step is generating a speckle pattern from an input azimuthal index. The
Eqn.~\ref{eqn:LGElecAss} is used to numerically generate the complex electric field of the input beam to the scattering media with $w_0\,{=}\,1$\,mm at a wavelength of 1064\,nm.
The complex electric field was evaluated over a $8\,{\times}\,8$\,mm area, centred on the beam, with discretisation of $20\,\mu$m resulting in a grid of $41\,{\times}\,41$ elements.
This size beam and sampling area was chosen as it roughly matches the beam size incident upon the scattering media in our experiment.
To simulate the effect of the scattering media, each element of the grid was assigned a random phase shift, to simulate the effect of a random thickness across the whole area of the media such as that used in the experiments described here.

Experimentally, the speckle pattern is observed in the far field. 
That is, the distance between the scattering media and camera is much greater than the wavelength of the light and the roughness of the scattering media, characterised by a Fresnel number much less than unity.
In the far field, the propagation of the speckle patterns can be described by Fraunhofer diffraction.

\begin{figure}[t]
    \centering
    \includegraphics[width=16cm]{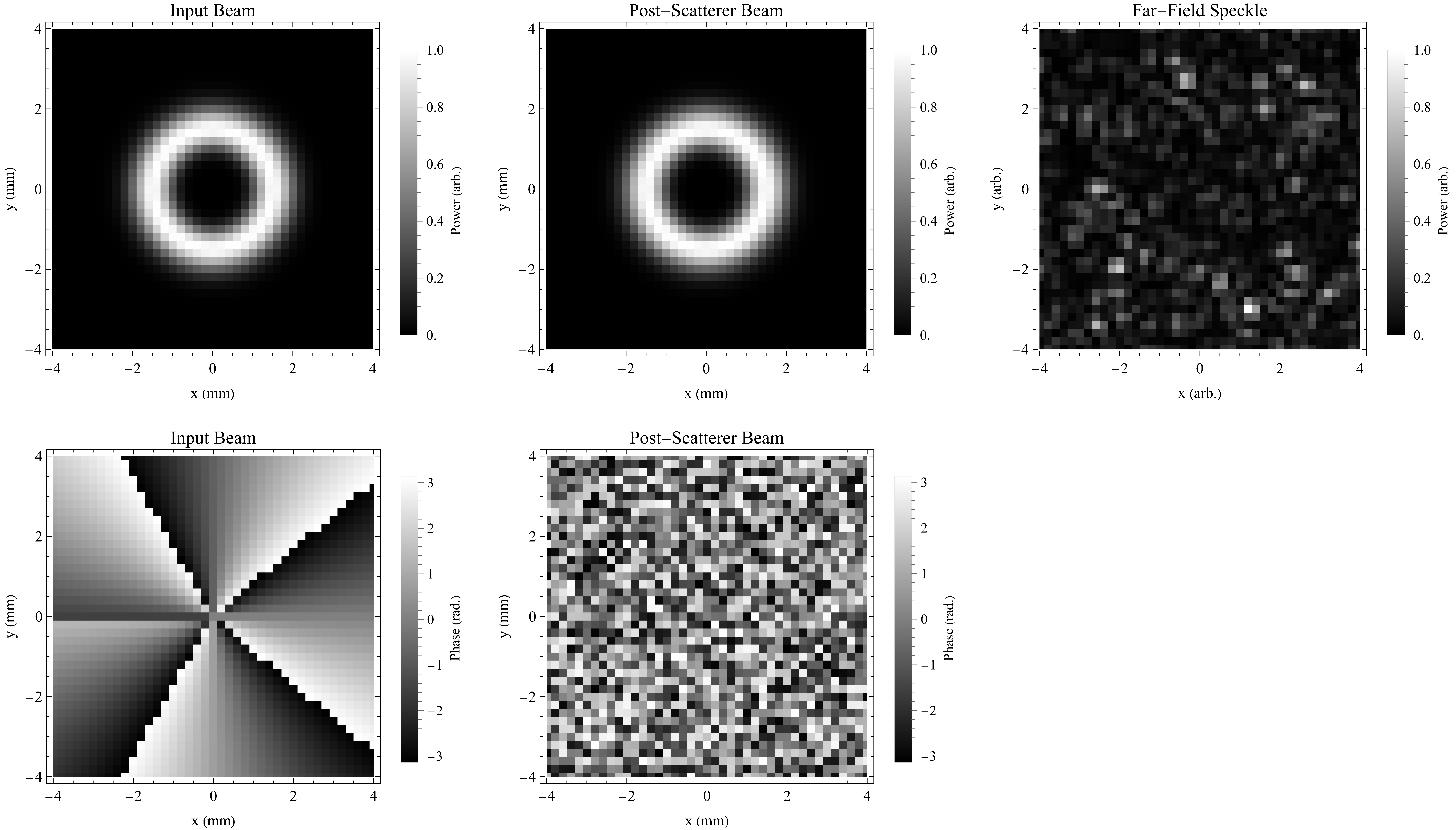}
    \caption{Simulated optical fields immediately before and after the scatterer (left and middle respectively), and in the far field showing a fully developed speckle pattern (right).
    The top row is the power of the optical field, while the bottom row shows the optical phase of the field.
    }
    \label{fig:SimSteps}
\end{figure}

Numerically, the speckle pattern is generated by implementing the Fraunhofer diffraction equation via a two-dimensional Fourier transform of the electric field after the random phase shift is imparted by the scattering media.
The result of this calculation is a speckle pattern generated over a grid of $41\,{\times}\,41$ elements.
Examples of the electric field, pre-scatterer, post-scatterer, and far field speckle are shown in Fig.~\ref{fig:SimSteps}.

To determine the output dependency of PCA upon input speckle patterns, speckle patterns with differing azimuthal indices were generated over the experimental range tested.
Two examples are shown here, for azimuthal indices which match what is experimentally observed for SPP2 (designed for $l\,{=}\,2$ when immersed in heavy water at 1064\,nm) and SPP5 (designed for $l\,{=}\,5$ when immersed in heavy water at 1064\,nm).
For SPP2, azimuthal indices between $l\,{=}\,1.84$ to 1.92 in steps of 0.01 were numerically generated.
For SPP2, azimuthal indices between $l\,{=}\,4.60$ to 4.78 in steps of 0.01 were numerically generated.
To analyse the speckle patterns using PCA, the numerically generated speckle patterns are processed in the same way that the experimental data is processed.
For details, see the Methods section of the main text.

\begin{figure}[t]
    \centering
    \includegraphics[width=5.2cm]{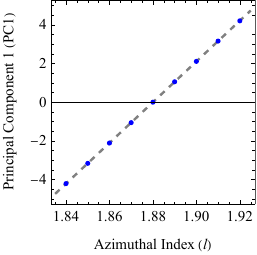}
    \quad
    \includegraphics[width=5.2cm]{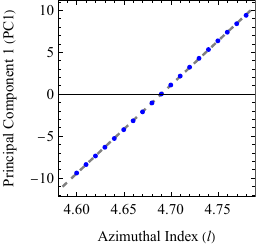}
    \quad
    \includegraphics[width=5.2cm]{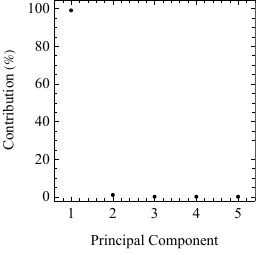}
    \caption{Principal component analysis (PCA) analysis of numerically generated speckle patterns.
    Left: First principal component (PC1) for PCA of speckle patterns arising from beams with azimuthal indices between $l\,{=}\,1.84$ to 1.92 in steps of 0.01.
    Middle: PC1 for PCA of speckle patterns arising from beams with azimuthal indices between $l\,{=}\,4.60$ to 4.78 in steps of 0.01.
    The PCA for both azimuthal index ranges shows a linear relationship between PC1 and azimuthal index.
    Right: The variance of the first 5 principal components, showing ${\approx}\,99\,\%$ of the variance of the data is captured in PC1. 
    This plot is the same for PCA on both azimuthal index ranges shown in the left and middle graphs.
    }
    \label{fig:PCANumerical}
\end{figure}

The result of the PCA analysis of the numerically generated speckle patterns is seen in Fig.~\ref{fig:PCANumerical}, for azimuthal indices matching SPP2 and SPP5 on the left and middle graphs, respectively.
For both azimuthal index ranges, the first principal component (PC1) contains ${\approx}\,99\,\%$ of the variance of the data and forms a near-linear relationship between PC1 and azimuthal index.
This agrees with what is observed experimentally, where PC1 contains ${\approx}\,90\,\%$ of the variance of the data, not as high as the numerical data which is due to experimental noise and drifts in the datasets.
This highlights that the PCA is only dependent upon the change in azimuthal index and other environmental changes are not relevant according to this analysis.
Similarly, the experimental data represented in PC1 linearly depends upon the azimuthal index as we would expect.

From this model, it can be seen that PCA of speckle patterns with differing azimuthal indices returns a linear relationship between PC1 and the azimuthal index, as shown in Fig.~\ref{fig:PCANumerical}.
The gradient of the linear relationship is the same for both azimuthal index ranges simulated, Fig.~\ref{fig:PCANumerical} left and middle.
This is a result of the sensitivity of speckle patterns to changes in the azimuthal index being independent of the magnitude of the azimuthal index of the beam.
This makes PCA a simple and powerful analysis technique for extracting the azimuthal index from speckle patterns.
Furthermore, the simple linear relationship between PC1 and azimuthal index allows for interpolation between training points to determine azimuthal indices that were not used in the training of the PCA transformation matrix to be measured with accuracy.

\section{Measurement Sensitivity} \label{sec:Measurement Sensitivity}
There are two experimental parameters that significantly affect the measurement sensitivity.
We define the measurement sensitivity as the change in azimuthal index induced by a change in the refractive index of the media being measured.
The larger the change in azimuthal index for a given refractive index change, the greater the measurement sensitivity.
For a given sample refractive index, $n$, a beam with azimuthal index, $l$, is generated which follows:
\begin{equation}
   l =  \left(n_\textrm{SPP} - n\right) \frac{h_s}{\lambda},
    \label{eqn:lvsn}
\end{equation}
where the $\lambda$ is the wavelength of light used, $h_s$ is the height of the spiral phase plate (SPP), and $n_\textrm{SPP}$ is the refractive index of the SPP material.
We quantify the measurement sensitivity as the factor $h_s/\lambda$, which depends upon the height of the SPP and laser wavelength.
This is a critical factor as the larger the change in azimuthal index generated by a change in refractive index, the larger the change in speckle pattern which is measured. 

In turn, the greater the measurement sensitivity we achieve with $l$, the more precisely we can measure changes in refractive index.
As shown in the main text, the refractive index of the samples, $n$, is related to the generated beam's azimuthal index, $l$, for which the speckle patterns are a proxy, by:
\begin{equation}
   n =  n_\textrm{SPP} - \frac{l\lambda}{h_s},
    \label{n vs l}
\end{equation}
which is a rearrangement of Eqn.~\ref{eqn:lvsn}.
The measurement precision of the refractive index, $\partial n$, is given by the partial differential of Eqn.~\ref{n vs l}, being:
\begin{equation}
   \partial n = -\frac{\lambda}{h_s}\partial l,
    \label{del n vs del l}
\end{equation} 
which quantifies the refractive index measurement noise, $\partial n$, as a consequence of the wavelength, SPP height, and azimuthal index measurement noise $\partial l$.
Section~\ref{sec:SpeckleAnalysis} showed that the sensitivity of the azimuthal index, $l$, was constant, independent of the absolute value of $l$ produced by the SPP.
That is to say, using a different height SPP or laser wavelength to change the magnitude of $l$ does not affect the sensitivity of the azimuthal index measurement.
Given experimental noise sources are constant, this implies that the noise of the azimuthal index measurement, $\partial{l}$, is also constant, which we observe in our measurements.
Thus, Eqn.~\ref{del n vs del l} shows that the noise of the refractive index measurements, $\partial{n}$, can only be changed by changing the wavelength and the height of the plates used in the experiment.
It can be seen that maximising the measurement sensitivity ($h_s/\lambda$) in Eqn.~\ref{eqn:lvsn}, also minimises the measurement noise ($\lambda/h_s$) in Eqn.~\ref{del n vs del l}.
Thus, a higher sensitivity leads to lower refractive index measurement noise and higher measurement precision. 
In the main text, we refer to the measurement resolution, which we define as the precision of the refractive index measurement.
The following sections discuss the effect of using different plate heights and different wavelengths.

\subsection{Effect of Plate Height}

\begin{figure}[t]
    \centering\includegraphics[width=12cm]{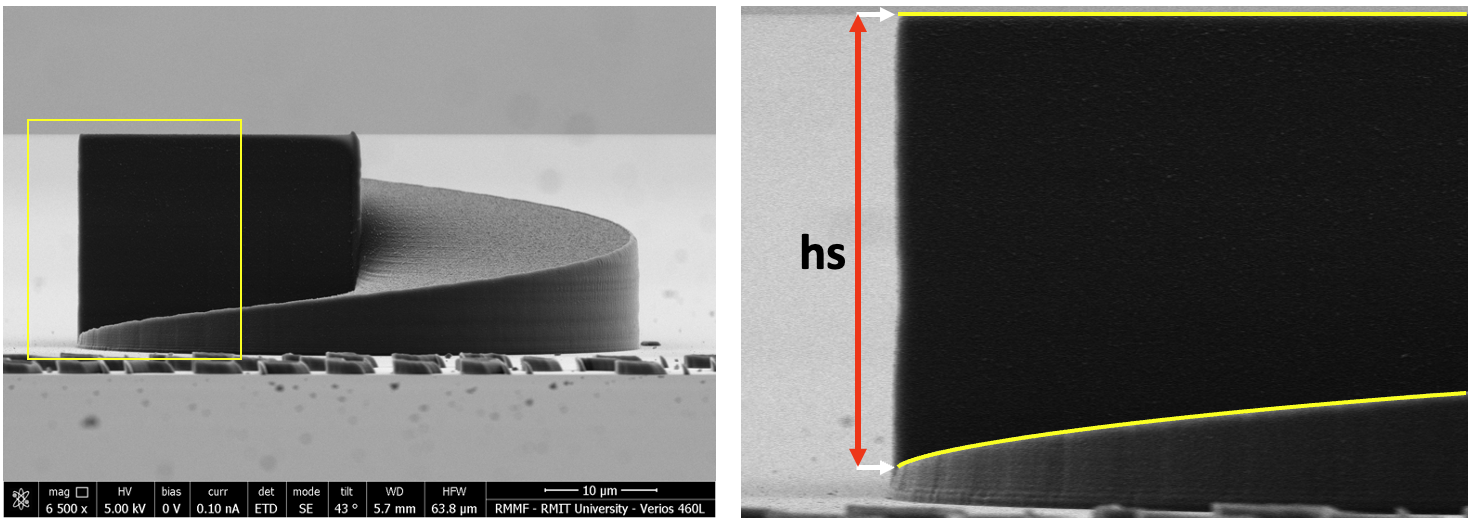}
    \caption{Scanning electron microscope image of the SPP used to measure error in height due to fabrication. 
    The SPP height, $h_S$, is marked with the red line.}
    \label{fig:SPPheight}
\end{figure}

We first consider the effect that the SPP height has upon the measurement.
The SPP `height' is a measure of the step height, $h_s$, that results from the inclined spiral as shown in Fig.~\ref{fig:SPPheight}.
The greater the plate height, $h_s$, and the accrued optical phase shift leading to a larger azimuthal index of the beam that is created by passing through the SPP, as described by Eqn.~\ref{eqn:lvsn}.

\begin{figure}[t]
    \centering
    \includegraphics[width=11cm]{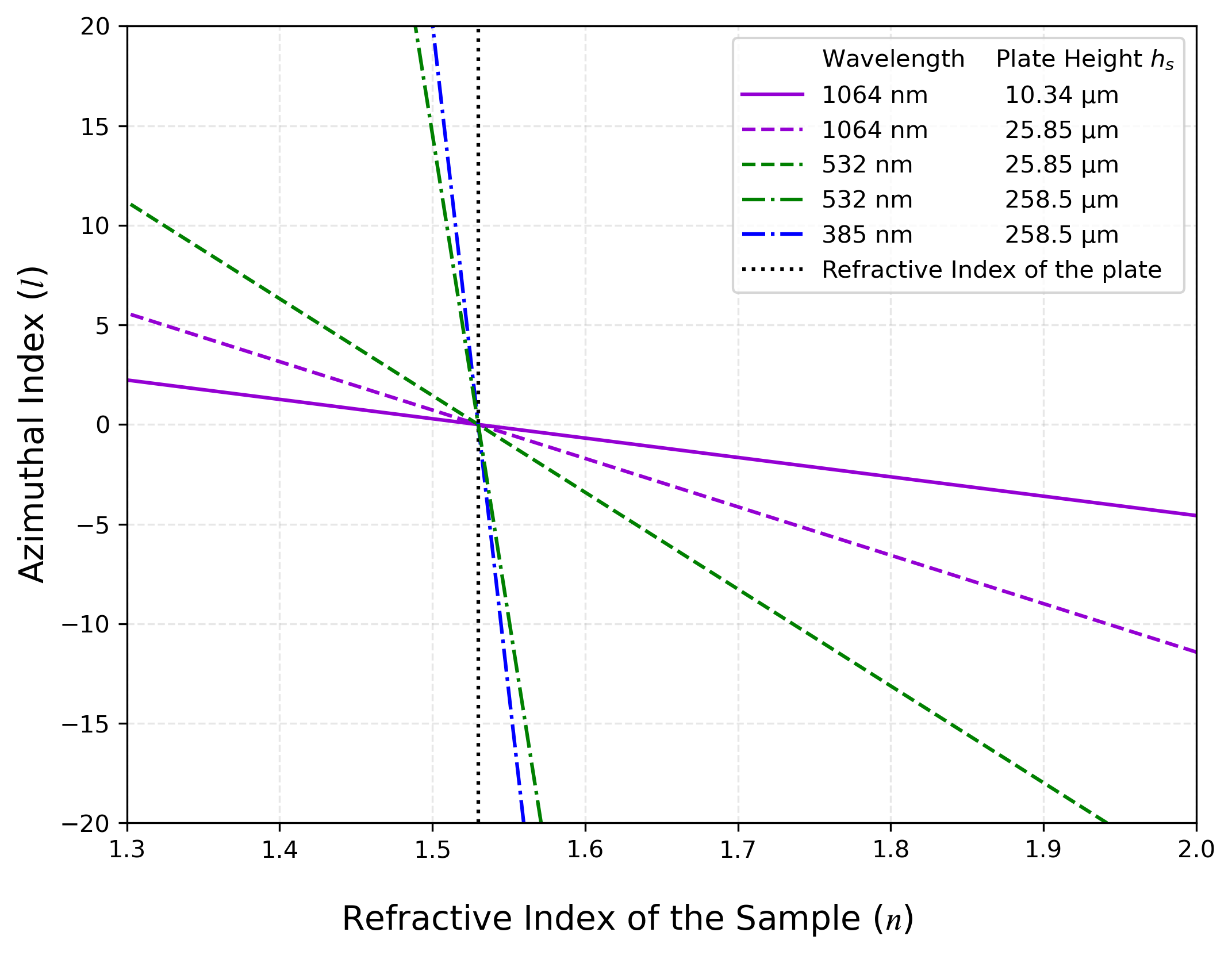}
    \caption{Plot of the relationship between azimuthal index and the refractive index of the surrounding for different step height and wavelength plotted using Eqn.~\ref{del n vs del l}.}
    \label{fig:Sensitivity}
\end{figure}

Figure~\ref{fig:Sensitivity} shows how the azimuthal index of the generated beam changes with sample refractive index, described by Eqn.~\ref{eqn:lvsn}.
Figure~\ref{fig:Sensitivity} also shows the effect of different SPP heights and laser wavelengths.
The gradient of each line, $h_s/\lambda$, is the measurement sensitivity. 
For better measurement sensitivity, a large change in azimuthal index needs to correspond to a small change in refractive index.
This can be achieved by increasing the height of the SPP, which is shown in Fig.~\ref{fig:Sensitivity} by a steeper gradient line.
An example is shown in Fig.~\ref{fig:Sensitivity} for a wavelength of $\lambda\,{=}\,1064$\,nm, comparing a plate height of $h_s\,{=}\,25.85\,\mu$m (purple dashed line) to a plate height of $h_s\,{=}\,10.34\,\mu$m (purple solid line), which shows a 2.5-fold increase in sensitivity (gradient).
A corresponding 2.5-fold reduction in measurement noise, Eqn.~\ref{del n vs del l}, and improvement in measurement precision, is also expected.
This example corresponds to SPP2 and SPP5 from the main text, with heights of $h_s\,{=}\,25.85\,\mu$m and $h_s\,{=}\,10.34\,\mu$m respectively, for which we observe a 2.5-fold increase in measurement precision when using SPP5 compared to SPP2.

For future studies, this could potentially be further improved if a plate with much greater height is used for the measurements. 
For example, a 10-fold improvement can be achieved if a plate with height $h_s\,{=}\,258.5\,\mu$m is used compared to a plate with $h_s\,{=}\,25.85\,\mu$m (purple dashed line) height and a 25-fold improvement compared to $h_s\,{=}\,10.34\,\mu$m (purple solid line).
We restrict ourselves to under 300$\mu$m plate height, as this is the height limit of the Nanoscribe fabrication facility used here.
This could point towards refractive index measurements pushing towards $7\,{\times}\,10^{-8}$, for a $h_s\,{=}\,258.5\,\mu$m plate height measured at 1064\,nm, assuming all measurement noise sources remain constant.

\subsection{Effect of Laser Wavelength}
Similar to the SPP height, an improvement in measurement sensitivity can be achieved when a shorter illumination wavelength is used.
The shorter the wavelength of the light, the larger the phase shift acquired when passing through a SPP of a given height, thus generating a larger azimuthal index, as described by Eqn.~\ref{eqn:lvsn}.
This in turn leads to a larger change in azimuthal index for a given change in refractive index.
A further consequence is a reduction in the noise of the refractive index measurement, $\partial n$, as described by Eqn.~\ref{del n vs del l}.

To illustrate this point, Fig.~\ref{fig:Sensitivity} compares lasers at wavelengths of 1064\,nm (purple lines) and 532\,nm (greenlines), showing a laser at 532\,nm has twice the measurement sensitivity (line gradient) of the 1064\,nm laser, as it is half the wavelength of 1064\,nm and thus accrues twice as much phase passing through the SPP. 
Thus, a wavelength of 532\,nm (green dashed line) will result in two-fold improvement in sensitivity and measurement precision when compared to using a wavelength of 1064\,nm (purple dashed line), with the plotted example being for a plate height of $h_s\,{=}\,25.85\,\mu$m.
This exact example corresponds to measurements of SPP5 in the main text, where we showed this two-fold improvement experimentally when comparing measurements taken at 1064\,nm and 532\,nm.

The sensitivity could be further improved by using a laser wavelength of 385\,nm (blue dash-dot line), instead of 1064\,nm, resulting in a 2.76-fold improvement in sensitivity and measurement precision.
Here, we restrict ourselves to 385\,nm as working with anything below this wavelength requires specialised optics and sensors. 

By combining changes to the laser wavelength and SPP height, an even larger increase in measurement sensitivity and precision can be achieved.
Two examples of this are shown in Fig.~\ref{fig:Sensitivity}.
The first example is using a laser wavelength of 532\,nm passing through a SPP with height of $h_s\,{=}\,258.5\,\mu$m (green dash-dot line) which will yield an improvement of 50-fold when compared to $h_s\,{=}\,10.34\,\mu$m and 1064\,nm wavelength (purple solid line).
A second example is to extend this further by using a 385\,nm laser wavelength through an SPP with a height of $h_s\,{=}\,258.5\,\mu$m (blue dash-dot line) that would improve the measurement sensitivity and precision 70-fold, compared to the 1064\,nm wavelength and $h_s\,{=}\,10.34\,\mu$m (purple solid line). 
This points towards refractive index measurements pushing towards $3\,{\times}\,10^{-8}$, assuming if all measurement noise sources remain constant.

\section{Measurement Accuracy}
The speckle measurement the azimuthal index of the laser beam, from which we infer the refractive index of the sample, are both measured against a training set of data.
Thus, the accuracy of a given measurement is limited by the accuracy of the training set of data.
In this section, we discuss the limits of the accuracy of the training set and thus influence the measurement accuracy of the system.

The accuracy can be broken down into two categories: those that affect the refractive index measurement and those that affect the azimuthal index measurement.
First looking at the refractive index measurement, the limit originates from the knowledge of the refractive index of the training samples.
It should be noted that the accuracy of the azimuthal index measurement does not impact the refractive index measurement.
Secondly, considering the azimuthal index of the generated beam, its accuracy depends upon knowledge of the refractive index, along with the height of the SPP and its refractive index at the laser wavelength used.

\subsection{Wavelength dependency of sample refractive index}
The refractive index of the sucrose solution we use for training the system is, as expected, wavelength dependant.
The refractive indices of our training samples were measured with a commercial refractometer, which are typically designed and calibrated to work at a particular wavelength.
The refractometer utilised in this work was a MyBrix, Mettler Toledo, which operates at the sodium line D-line (Na$_D$) of $\approx\,589.29$\,nm.

The speckle measurements are performed at laser wavelengths of 532\,nm and 1064\,nm, at which the training samples will present different refractive indices from that measured by the commercial refractometer.
To quantify the change in refractive index at different wavelengths, we use the Sellmeier equation,  that describe the wavelength dependence of the refractive index of a transparent medium and wavelength~\cite{ghosh1997sellmeier, ghosh1994temperature, bhar1980temperature}.
Though other similar empirical approaches exist, such as the Cauchy equation, which also describe the wavelength dependence of refractive index, we utilise the Sellmeier equation.
We deem this to be a more accurate model over a much broader spectrum, often covering the ultraviolet, visible, and infrared spectrum~\cite{ghosh1997sellmeier, ghosh1994temperature, bhar1980temperature}.

For water, which is the $0\,\%$ sucrose solution, the coefficients for the Sellmeier equation have previously been measured~\cite{daimon2007measurement}:
\begin{equation} \label{eqn:sellmeiers_water}
   n^{2}(\lambda) = 1 + \frac{0.5689\lambda^{2}}{\lambda^{2} - 5.11\times 10^{3}} + \frac{0.1719\lambda^{2}}{\lambda^{2} - 1.825\times 10^{4}} + \frac{0.0206\lambda^{2}}{\lambda^{2} - 2.624\times 10^{4}} + \frac{0.1124\lambda^{2}}{\lambda^{2} - 10.6750\times 10^{6}},
\end{equation}
for wavelengths, $\lambda$, measured in nano-meters, nm. 
The equation was determined over the wavelength range 182\,nm - 1129\,nm, covering the wavelengths used here.

Similarly for a $5\,\%$ sucrose solution, the Sellmeier coefficients have been previously measured~\cite{purwandari2021analyses}: 
\begin{equation} \label{eqn:sellmeiers 5}
   n^{2}(\lambda) = 1 + \frac{0.788\lambda^{2}}{\lambda^{2} - 578.714},
\end{equation}
for wavelengths, $\lambda$, measured in nano-meters, nm.
The equation was determined over the wavelength range 455\,nm - 633\,nm.

These equations will differ for different sucrose concentrations. 
As the working range in our measurements (for our work being at 532\,nm and 1064\,nm) was between 0-5\% of sucrose concentrate, we only consider the 0\% and 5\% concentration for easy comparison. 
For any solutions between 0\% to 5\%, the Sellmeier equations could not be found in the literature, so the comparison could be drawn only for 0\% and 5\% solutions.

\begin{table}[ht] 
\centering
\begin{tabular}{|c|c|c|c|}
\hline
\multicolumn{1}{|l|}{Wavelength (nm)} & \begin{tabular}[c]{@{}c@{}}Refractive index  \\ from Eqn.~\ref{eqn:sellmeiers_water}\end{tabular} & \begin{tabular}[c]{@{}c@{}}Measured \\ refractive index\end{tabular} & \begin{tabular}[c]{@{}c@{}}Difference \\ in refractive index\end{tabular} \\ \hline
532                                   & 1.3352                                                            & --                                                                   & 0.180\,\%                                                                   \\ \hline
589.29                                & 1.3332                                                            & 1.3328                                                               & 0.030\,\%                                                                    \\ \hline
1064                                  & 1.3244                                                            & --                                                                   & -0.630\,\%                                                                   \\ \hline
\end{tabular}
\caption{Table showing refractive index of 0\,\% sucrose solution determined by 
    Sellmeier equation from Eqn.~\ref{eqn:sellmeiers_water} at different wavelengths.
    It also shows the percentage difference between the refractive index determined by the empirical equations and the refractive index measured by the refractometer for the 0\,\% sucrose solution at a wavelength of 589.29\,nm.
    }
\label{tab:0percenttable}
\end{table}

\begin{table}[ht]
\centering
\begin{tabular}{|c|c|c|c|}
\hline
\multicolumn{1}{|l|}{Wavelength (nm)} & \begin{tabular}[c]{@{}c@{}}Refractive index  \\ from Eqn.~\ref{eqn:sellmeiers 5}\end{tabular} & \begin{tabular}[c]{@{}c@{}}Measured \\ refractive index\end{tabular} & \begin{tabular}[c]{@{}c@{}}Difference \\ in refractive index\end{tabular} \\ \hline
532                                   & 1.3377                                                            & --                                                                   & -0.189\,\%                                                                   \\ \hline
589.29                                & 1.3376                                                            & 1.3403                                                               & -0.197\,\%                                                                   \\ \hline
1064                                  & out of range of fitting                                           & --                                                                   &                                                                           \\ \hline
\end{tabular}
\caption{Table showing refractive index of 5\,\% sucrose solution determined by 
    Sellmeier equation from Eqn.~\ref{eqn:sellmeiers 5} at different wavelengths.
    It also shows the percentage difference between the refractive index determined by the empirical equations and the refractive index measured by the refractometer for the 5\,\% sucrose solution at a wavelength of 589.29\,nm.
    }
\label{tab:5percenttable}    
\end{table}

Table~\ref{tab:0percenttable} and ~\ref{tab:5percenttable} compare the refractive indices measured from the refractometer (at the sodium D-line) for 0\,\% and 5\,\% sucrose solution respectively, and compare this to the refractive indices expected at the wavelengths used in the speckle measurements, calculated by the Sellmeier equation given by Eqn.~\ref{eqn:sellmeiers_water} and Eqn.~\ref{eqn:sellmeiers 5} respectively. 
The tables show a maximum error of 0.63\,\%, which is at 1064\,nm for the 0\,\% sucrose solution.
Unfortunately, the characterisation of refractive index for a 5\% sucrose solution, given by  Eqn.~\ref{eqn:sellmeiers 5}, did not include 1064\,nm.
It is expected that the error for a 5\,\% sucrose solution at 1064\,nm, is similar to that of the 0\,\% sucrose solution. 
This is justified by noting that the difference between the refractive indices at 0\,\% and 5\,\%, both for the calculated refractive indices using Eqn.~\ref{eqn:sellmeiers_water} and Eqn.~\ref{eqn:sellmeiers 5}, along with the measured refractive indices, are all below the maximum error of 0.63\,\% for 1064\,nm in 0\,\% solution.
Furthermore, a Sellmeier equation for intermediary concentrations of sucrose solution could not be found, thus, the precise errors for other sucrose solutions cannot be estimated. 
This also means we cannot take these refractive index offsets into account for the training solutions at the working wavelength.

This comparison reveals that the maximum error of 0.63\% (a difference of about $84\,{\times}\,10^{-4}$\,RIU) can be expected due to the wavelength dependency of the refractive index of the sucrose solution. 
This is a significant difference when we compare this with the accuracy of the commercial refractometer, which is about 0.0225\,\% (the refractometer has an accuracy of $3\,{\times}\,10^{-4}$\,RIU).

\subsection{Wavelength dependency of the spiral phase plate}
The SPP used in the experiment is fabricated with IP-DIP resin, which has a refractive index, $n_\textrm{SPP}$, that is also wavelength dependent, giving slightly different values at 1064\,nm and 532\,nm.
Previous works have experimentally measured the refractive index of IP-DIP at different wavelengths~\cite{schmid2019optical}, giving the coefficients for the Sellmeier equation to give: 
\begin{equation}    \label{eqn:sellmeiersIP-DIP}
    n_\textrm{SPP}^{2}(\lambda) = 1 + \frac{1.2899\lambda^{2}}{\lambda^{2} - 1.1283\times 10^{4}} + \frac{0.060569\lambda^{2}}{\lambda^{2} - 7.7762\times 10^{4}} + \frac{118440\lambda^{2}}{\lambda^{2} - 2.5802\times 10^{13}}
\end{equation}
for wavelengths, $\lambda$, measured in nano-meters, nm. 
The equation was determined over the wavelength range 405\,nm - 1400\,nm, covering the wavelengths used here.

Eqn.~\ref{eqn:sellmeiersIP-DIP} is used to estimate of refractive index of the plates at both 532\,nm and 1064\,nm. The
Eqn.~\ref{eqn:sellmeiersIP-DIP} gives the plates refractive index at 1064\,nm to be $n_\textrm{SPP}(1064\,\textrm{nm})\,{=}\,1.53$.
This value was used in designing the SPPs, and for calculating the azimuthal index of the generated beams when working at 1064\,nm.
At a wavelength of 532\,nm, the refractive index of the plate increases by $1.3\,\%$ to $n_\textrm{SPP}(532\,\textrm{nm})\,{=}\,1.55$, this value was used for calculating the generated beams azimuthal index when working at a wavelength of 532\,nm.
As we use these values in our calculations, this eliminates any error associated with the wavelength dependency of the refractive index for IP-DIP resin.

\subsection{Temperature dependency of refractive index} \label{sec:TempRefIndex}
The refractive index of the sucrose solution, haemoglobin, and SPP are not only dependent on the laser wavelength but also on their respective temperatures.
The experiments performed here were not temperature-controlled; thus any change in the temperature of the sample results in a change in the refractive index.
Over the time scale of the whole measurement run (training and precision measurements) which was ${\approx}\,30$\,min, the temperature, $T$, of the room and sample changed by approximately ${\pm}\,50$\,mK.
While over the time scale of a precision measurement, typically ${\approx}\,30$\,sec, the room and sample changed by approximately ${\pm}\,5$\,mK.
These temperature fluctuations were measured using a 10\,k$\Omega$ thermistor attached to the microfluidic chamber with a Kapton tape.
To quantify the change in refractive index caused by this change in temperature, we use the material's thermo-optic coefficient ($\frac{dn}{dT}$). 
Previous work has determined the temperature dependency of the refractive index of water~\cite{daimon2007measurement} to be given by the equation:  
\begin{equation}    \label{eqn:dn/dt water}
    \frac{dn}{dT} = -8.47 \times 10^{-5},
\end{equation}
quantified at a wavelength of 1083.33\,nm, and in units of RIU/Kelvin.
From this, we calculate the change in refractive index over the whole measurement run to be ${\pm}\,4.2\,{\times}\,10^{-6}$\,RIU.

Similarly, for haemoglobin, the temperature dependence of its refractive index was previously determined~\cite{lazareva2016temperature}:
\begin{equation}    \label{eqn:dn/dt haemoglobin}
    \frac{dn}{dT} = -8.26 \times 10^{-5},
\end{equation}
quantified at a wavelength of 1100\,nm, and concentration of 1.25\,mmol/L. 
From this we calculate the change in refractive index over the whole measurement run to be ${\pm}\,4.1\,{\times}\,10^{-6}$\,RIU.

The thermo-optic coefficient for the sucrose solution and the SPP could not be found in the literature, so we limit our discussion to only water and haemoglobin. 
However, a similar change in refractive index (approximately ${\pm}\,4\,{\times}\,10^{-6}$\,RIU) can be expected for these samples.

Over the duration of the whole measurement run, the maximum change in refractive index associated with temperature changes is ${\pm}\,4.2\,{\times}\,10^{-6}$\,RIU which sets a limit on the accuracy of the measurement.
This corresponds to a change of $3.2\,{\times}\,10^{-4}$\%.
A temperature stabilised sample would overcome this limitation.
Over the duration of a precision measurement, ${\approx}\,30$\,sec, the change in refractive index is ${\pm}\,4.2\,{\times}\,10^{-7}$\,RIU, a change of $3.2\,{\times}\,10^{-5}$\%. 
This change is an order of magnitude lower than our measurement precision and thus can be neglected, see Sec.~\ref{sec:noise}.


\subsection{Fabrication accuracy of the spiral phase plate}
Fabrication of the SPPs affects the measurement accuracy as the plates will not have identical heights to one another, nor will they necessarily be exactly the designed height.
To quantify this error, we printed four sets of SPPs on different substrates, with each set consisting of five different plate heights, $h_s$, thus 20 plates in total.
These SPPs were imaged with a scanning electron microscopy (SEM) in the following manner: first, they were coated with 8\,nm Au/Pd, then the substrate was cleaved close to the plate set, the SPPs were mounted in the SEM at $45^\circ$, and tilted to an imaging angle of $88.5^\circ$.
This rotation was normal to the phase ramp step, and focus was on the bottom of the spiral ramp.
The height of the SPPs, $h_s$, was measured from the bottom of the ramp to the top of the ramp, as seen in Fig.~\ref{fig:SPPheight}.

The overall average offset from the design specifications for all 20 plates across all four sets was $-42\,{\pm}\,18$\,nm, with the error being the standard deviation across the 20 plates measured. 
For SPP2, this offset is 0.407\,\% of the plate height of $h_s\,{=}\,10.34\,\mu$m.
For SPP5, this fractional offset reduces to 0.162\,\% on the plate height of $h_s\,{=}\,25.85\,\mu$m. 
The fractional offset reduces for increasing plate height, thus improving the accuracy of our measurements.

This offset in between the designed and fabricated SPP height leads to an inaccuracy in calculating the azimuthal index of the generated beam.
There is a linear relationship between azimuthal index and plate refractive index, see Eqn.~\ref{eqn:lvsn}, thus for SSP2, the inaccuracy in the azimuthal index is 0.407\%, and for SSP5 it is 0.162\%.
This inaccuracy only impacts the measurement of $l$, and has no effect on the precision of our measurements. 
In future, this offset can be accounted for by accurate measurement of the plate height before use in the speckle experiment.

\subsection{Thermal expansion of the spiral phase plate}
Just as the accuracy of the fabrication affects the height of the SPP, the temperature also affects the height of the plates due to thermal expansion or contraction
Thus, changes in temperature affect the height of the SPP, and the azimuthal index of the generated beam, see Eqn~\ref{eqn:lvsn}. 
The linear thermal expansion coefficient, $\alpha_{L}$, for IP-DIP (the resin from which the plates were fabricated) characterises the fractional length change of the material, of length $L$, per unit change in temperature $\left(\frac{1}{L}\frac{dL}{dT}\right)$.
Previous work provides the linear thermal expansion coefficient for IP-DIP resin~\cite{qu2017micro} which is given by the following equation:
\begin{equation}    \label{eqn:thermal expansion IP_DIP}
    \alpha_{L} = \frac{1}{L}\frac{dL}{dT}=(5 \pm 0.5) \times 10^{-5}.
\end{equation}
Over the time scale of the whole measurement run, the sample changed temperature by ${\pm}\,50$\,mK, see Sec.~\ref{sec:TempRefIndex}. 
For SPP5, this corresponds to a height change of ${\pm}\,0.065$\,nm, or $2.5\,{\times}\,10^{-4}$\%.
This results in a change in the azimuthal index measurements of $2.5\,{\times}\,10^{-4}$\%, corresponding to a change of $1.2\,{\times}\,10^{-5}$.
For the refractive index measurement, a change of $5\,{\times}\,10^{-7}$\,RIU would be observed, or $4\,{\times}\,10^{-5}$\%.
This would also impact the accuracy to the measurement, again a temperature stabilised sample would overcome this limitation.

For the precision measurements, with a time scale of ${\approx}\,30$\,sec, the sample changed temperature by approximately ${\pm}\,5$\,mK.
This corresponds to a change in the azimuthal and refractive index measurements of $1.2\,{\times}\,10^{-6}$ and  $5\,{\times}\,10^{-8}$\,RIU.
This change is an order of magnitude lower than our measurement precision and thus can be neglected, see Sec.~\ref{sec:noise}. 

\section{Measurement precision - noise sources} \label{sec:noise}
In this section, the major sources of noise in the experiment are discussed and the effect that they have on the measurement precision. 
All sources of noise discussed here impact both the refractive index and azimuthal index measurements.
The major sources of noise identified are: SPP mechanical and beam-pointing instability, laser wavelength instabilities, shot noise, and camera noise.
The following subsection looks at the individual noise sources and their effect.
For each noise source, the measurement procedure is similar: the system is trained as described in the main text, following which the noise source being measured is investigated, as described below.

\subsection{Laser wavelength instability}

\begin{figure}[t]
    \centering
    \includegraphics[width=8cm]{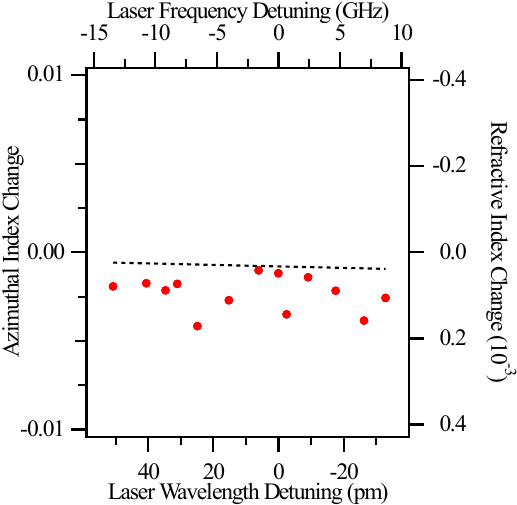}
    \caption{Wavelength/frequency dependence of the refractive index and azimuthal index measurements.
    The data was taken at a central wavelength of $\lambda\,{=}\,1064.4156$\,nm, with SPP5.
    The change in azimuthal index and refractive index measurements are deviations around central values of $l\,{=}\,4.7$ and $n\,{=}\,1.3365$, respectively.
    The data shows little dependence upon wavelength/frequency of the laser.
    The dashed line indicates the change in the measured values expected from a change in wavelength, which in turn changes the azimuthal index of the generated beam after the SPP, as expected from Eqn.~\ref{eqn:lvsn}.
    }
    \label{fig:WaveSens}
\end{figure}

Speckle patterns are known for their high sensitivity to changes in laser wavelength~\cite{bruce2019overcoming}.
In some instances this is advantageous, such as speckle-based wavemeters or spectrometers.
Here, however, speckle wavelength sensitivity introduces a source of noise to the measurement of a sample's refractive index or azimuthal index.
Here, a single scatterer is used to produce the speckle pattern to minimise sensitivity to wavelength changes, whilst maintaining sensitivity to azimuthal index changes~\cite{facchin2024determining}.
Whilst using a single scatterer reduces sensitivity to wavelength changes, the speckle pattern is still sensitive to this, which will impact our measurements.

To measure the wavelength sensitivities of the speckle measurement, an intentional laser wavelength change is made in our system which we analyse.
A second frequency tunable 1064\,nm laser (Coherent Mephisto) was used for this test. 
The laser frequency was changed by $\pm\,15$\,GHz, which was measured independantly by a commercial wavemeter (HighFinesse WS6-600).
Figure~\ref{fig:WaveSens} shows that there is general agreement between the data and the expected values, however noise dominates the measurement as such small changes are expected.
Changing the laser wavelength does not create a change in the speckle pattern measurement outside of what is expected from phase changes and beam shape changes induced by the SPP.

The wavelength stability of the 1064\,nm laser used for our measurements was also measured using the commercial wavemeter.
The time scale of a precision measurement was typically ${\approx}\,30$\,sec.
Over this duration, the wavelength stability of the laser is ${\approx}\,1$\,MHz, corresponding to change in the refractive index measurement of ${\approx}\,7\,{\times}\,10^{-10}$\,RIU.
Extending the time scale to that of a whole measurement run, that is including training and precision measurements (${\approx}\,30$\,min), the wavelength stability of the laser is ${\approx}\,20$\,MHz, corresponding to a change in the refractive index measurement of ${\approx}\,10^{-8}$\,RIU.
The laser stability over both timescales of interest is orders of magnitude lower than that used to test the sensitivity of the speckle pattern which showed little correlation.
Thus, this indicates that the effect of laser wavelength stability upon both the refractive index and azimuthal index measurements is below the current noise floor of the experiment.

\subsection{Optical alignment of the spiral phase plate}

\begin{figure}[t]
    \centering\includegraphics[width=8cm]{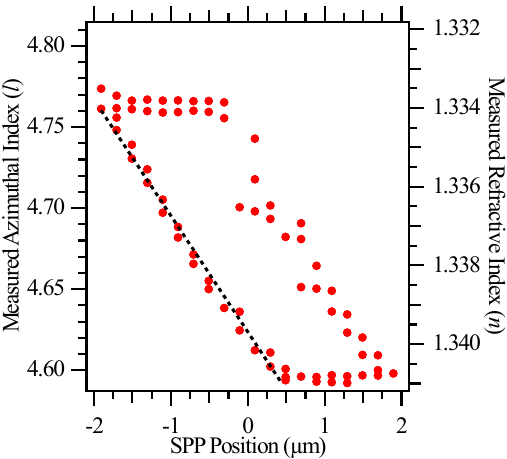}
    \caption{Position dependence of the refractive index and azimuthal index measurements for SPP5.
    The data clearly shows the hysteresis in the stepper motor as it was stepped forward and backwards around the mean position.
    Dotted line is a fit to the data moving in the forward direction giving a sensitivity of ${\approx}\,3\,{\times}\,10^{-3}$\,RIU/$\mu$m or an azimuthal index change of ${\approx}\,-0.07$/$\mu$m.}
    \label{fig:PosSens}
\end{figure}

Changes in the position of the SPP relative to the laser beam passing through it may influence the spatial profile of the beam after the SPP, and hence the speckle pattern it produces~\cite{Mazilu2012}.
Such changes in relative position can be due to the SPP shifting in its spatial location due to mechanical vibrations or drifts over time, or changes in the laser beam position, again due to vibrations or drifts over time.
As it is the relative position between the SPP and laser beam that is of importance, it implies that the mechanical stability of the SPP and beam pointing stability generate identical effects upon the spatial mode of the generated beam and speckle pattern.

To measure the effect of the relative position between the SPP and laser beam, we systematically displace the SPP and measure the effect upon the speckle pattern and resulting refractive index and azimuthal index measurements.
The SPP is mounted on a translation stage with a stepper-motor controlled actuator (Physik Instrumente M-230.25).
The system is trained at 1064\,nm with SPP5, which is held stationary.
After training, one test solution is flushed into the micro-fluidic system and kept there for the duration of the displacement study.
The SPP is moved in 200\,nm increments, over $\pm\,2\,\mu$m, with speckle measurements taken at each step.
The standard speckle analysis used to determine the effect of the displacements upon the refractive index and azimuthal index which are presented in Fig.~\ref{fig:PosSens} which shows a strong correlation between SPP displacement and calculated refractive index and azimuthal index.
The sensitivity to movement of the SPP is shown by the dotted line on Fig.~\ref{fig:PosSens} which gives a refractive index sensitivity of ${\approx}\,3\,{\times}\,10^{-3}$\,RIU/$\mu$m or an azimuthal index change of ${\approx}\,-0.07$/$\mu$m.
Thus, for a 200\,nm displacement of the SPP, which corresponds to 0.4\,\% of the 50\,$\mu$m plate diameter, we see a change of ${\approx}\,6\,{\times}\,10^{-4}$\,RIU and ${\approx}\,0.014$ for the refractive index and azimuthal index respectively, which corresponds to a fractional change of $0.04\,\%$ and $-0.3\,\%$.

At 1064\,nm, the average refractive index measurement precision for SPP5 is $(3.4\,{\pm}\,0.9)\,{\times}10^{-6}$\,RIU which corresponds to a relative position precision between the SPP and laser beam of ${\approx}\,1$\,nm.
This alignment precision between the SPP and laser beam is extremely stringent and required removal of any mechanically unstable optics components to reach this level.
We have not been able to measure the relative position between the SPP and laser beam to a precision nearing ${\approx}\,1$\,nm, and thus cannot put an estimate on the magnitude of this noise floor in our refractive index and azimuthal index measurements.
If this relative position precision between the SPP and laser beam were to be an issue, in future this noise floor may be reduced by a monolithic design which holds both the optics and SPP on one solid block which would constrain the relative position of the SPP and laser to below 1\,nm.
An alternative would be to utilise a SPP fabricated onto the end of an optical fibre~\cite{weber2017single}.

\begin{table}[ht]
    \centering
    \begin{tabular}{c | cc}
        Measurement         & Measured Refractive Index Precision                       & Shot Noise Contribution                                   \\
        \hline
        SPP2 at 1064\,nm    & $\left(7.4\,{\pm}\,0.5\right)\,{\times}\,10^{-6}$\,RIU    & $\left(6.0\,{\pm}\,1.0\right)\,{\times}\,10^{-6}$\,RIU    \\
        SPP5 at 1064\,nm    & $\left(3.4\,{\pm}\,0.4\right)\,{\times}\,10^{-6}$\,RIU    & $\left(2.4\,{\pm}\,0.5\right)\,{\times}\,10^{-6}$\,RIU    \\
        SPP5 at 532\,nm     & $\left(2.2\,{\pm}\,0.3\right)\,{\times}\,10^{-6}$\,RIU    & $\left(1.2\,{\pm}\,0.2\right)\,{\times}\,10^{-6}$\,RIU    \\
    \end{tabular}
    \caption{Summary of the refractive index measurement precision described in the main text, compared to the estimated shot noise contribution.}
    \label{tab:ShotNoiseRefIndex}
\end{table}

\subsection{Shot Noise}

\begin{table}[t]
    \centering
    \begin{tabular}{c | cc}
        Measurement         & Measured Azimuthal Index Precision                       & Shot Noise Contribution                        \\
        \hline
        SPP2 at 1064\,nm    & $\left(7.2\,{\pm}\,0.5\right)\,{\times}\,10^{-5}$    & $\left(5.8\,{\pm}\,0.9\right)\,{\times}\,10^{-5}$  \\
        SPP5 at 1064\,nm    & $\left(8.6\,{\pm}\,1.0\right)\,{\times}\,10^{-5}$    & $\left(5.9\,{\pm}\,1.1\right)\,{\times}\,10^{-5}$  \\
        SPP5 at 532\,nm     & $\left(10.6\,{\pm}\,1.7\right)\,{\times}\,10^{-5}$   & $\left(6.0\,{\pm}\,0.8\right)\,{\times}\,10^{-5}$  \\
    \end{tabular}
    \caption{Summary of the azimuthal index measurement precision described in the main text, compared to the estimated shot noise contribution.}
    \label{tab:ShotNoiseAziIndex}
\end{table}

Noise associated with the random arrival of photons at the camera, termed shot noise, is the dominant noise source in the refractive index and azimuthal index measurements.
Shot noise manifests differently for each speckle image, that is the shot noise on each pixel is different due to the amount of light falling on it, and from frame to frame due to photon statistics, leading to noise in the final refractive index and azimuthal index measurement.

To measure the effect of shot noise, we train the system, then take multiple images with a shot noise limited light source shone through the experiment in place of the 1064\,nm or 532\,nm laser, which uniformly illuminates the camera.
These images are processed using the same method as the precision measurement images.
The light source was confirmed to be shot noise limited by changing the illuminating power of the light source, $P$, and measuring the standard deviation of the pixel values which showed the expected $\sqrt{P}$ behaviour for a shot noise limited light source.

Table~\ref{tab:ShotNoiseRefIndex} summarises the refractive index measurement precision, described in the main text, compared to the estimated shot noise contribution.
While Table~\ref{tab:ShotNoiseAziIndex} summarises the azimuthal index measurement precision, described in the main text, comparing them to the estimated shot noise contribution.
Both the measured refractive index and azimuthal index precision is consistent with the shot noise measurements, strongly suggesting the precision measurements are limited by photon shot-noise.

\subsection{Camera Noise}
Noise associated with the camera (CMOS Alvium 1800 u-158m, Allied Vision) feeds into the final precision of the refractive index and azimuthal index measurements.
Here we refer to camera noise as the combination of read noise and dark current noise that will be present when each pixel is read out~\cite{Peterkovic2025}.
For each speckle image, the camera noise on each pixel is different due to the statistics of the read and dark current noise, leading to noise in the final refractive index and azimuthal index measurement.

To measure the effect of camera noise, we train the system, then take multiple images with the laser turned off, resulting in dark images on the camera.
The dark images are processed using the same method as the bright precision measurement images.
The result shows a contribution of approximately $10^{-8}$ and $10^{-10}$ for the refractive index and azimuthal index measurements respectively.
This indicates the noise contribution from camera noise are many orders of magnitude below the demonstrated measurement precision, and thus are negligible.

